\documentstyle[12pt]{article}

\renewcommand{\bar}[1]{\overline{#1}}
\newcommand{\M}{{\cal M}}
\newcommand{\VEV}[1]{\left\langle{#1}\right\rangle}
\newcommand{\etal}{{\em et al.}}
\newcommand{\ie}{{\it i.e.}}
\newcommand{\eg}{{\it e.g.}}

\newcommand{\longvec}[1]{\overrightarrow{\!\!#1}}

\newcommand{\ket}[1]{\vert\,{#1}\rangle}
\newcommand{\ms}{$\overline{\mbox{MS}}$}

\newcommand{\amst}{\mbox{${\widetilde{\alpha}_{\overline{\mbox{\tiny MS}}}}$}}
\newcommand{\av}{\mbox{$\alpha_{V}$}}

\renewcommand{\bar}{\overline}

\newcommand{\GeV}{{\rm GeV}}

\newcommand{\cf}{{\em c.f.}}
\newcommand{\eqcm}{\: ,}
\newcommand{\lqcd}{\Lambda_{QCD}}
\newcommand{\dpt}{\Delta_\perp}
\input{epsf}

\begin{document}
\begin{flushright}
SLAC--PUB--8240\\
September 1999
\end{flushright}
\bigskip\bigskip

\thispagestyle{empty}
\flushbottom

\begin{center}
{{\Large \bf
QCD Technology:\\
Light-Cone Quantization and Commensurate Scale Relations}\footnote{\baselineskip=14pt
     Work supported by the Department of Energy, contract
     DE--AC03--76SF00515.}}\\
\vspace{1.0cm}
Stanley J. Brodsky\\
{\it{Stanford Linear Accelerator Center\\
Stanford University,
Stanford, California 94309}}\\
e-mail: sjbth@slac.stanford.edu\\
\end{center}

\vfill
\begin{center}
Lectures presented at the\\
The Twelfth Nuclear Physics Summer School and Symposium (NuSS'99)\\
and Eleventh International Light-Cone School and Workshop\\
May 26 - June 18, 1999, APCTP, Seoul, Korea
\end{center}
\vfill
\newpage

\begin{center}
Abstract
\end{center}

I discuss several theoretical tools which are useful
for analyzing perturbative and non-perturbative problems in quantum
chromodynamics, including (a) the light-cone Fock
expansion, (b) the effective charge $\alpha_V$, (c) conformal symmetry,
and (d) commensurate scale relations.  Light-cone Fock-state
wavefunctions encode the properties of a hadron in terms of its
fundamental quark and gluon degrees of freedom. Given the proton's
light-cone wavefunctions, one can compute not only the quark and gluon
distributions measured in deep inelastic lepton-proton scattering, but
also the multi-parton correlations which control the distribution of
particles in the proton fragmentation region and dynamical higher twist
effects. Light-cone wavefunctions also provide a systematic framework for
evaluating exclusive hadronic matrix elements, including timelike heavy
hadron decay amplitudes and form factors. The $\alpha_V$ coupling, defined
from the QCD heavy quark potential, provides a physical expansion
parameter for perturbative QCD with an analytic dependence on the fermion
masses which is now known to two-loop order. Conformal symmetry provides a
template for QCD predictions, including relations between observables
which are present even in a theory which is not scale invariant.
Commensurate scale relations are perturbative QCD predictions based on
conformal symmetry relating observable to observable at fixed relative
scale. Such relations have no renormalization scale or scheme ambiguity.
\bigskip

\section{Introduction}

A primary goal of both high energy and nuclear physics is to unravel the
structure and dynamics
of nucleons and nuclei in terms of their fundamental quark and gluon
degrees of freedom.
Our present empirical knowledge of the quark and gluon distributions of
the proton has revealed
a remarkably complex substructure.  It is helpful to categorize the parton
distributions as
``intrinsic" --pertaining to the composition of the target hadron, and
``extrinsic", reflecting
the  substructure of the individual quarks and gluons
themselves.  For example, the
$\bar u(x)$ and
$\bar d(x)$ antiquark distributions of the proton at $Q^2
\sim 10$ GeV$^2$ to be quite different in shape\cite{{Nasalski:1994bh}} and
thus must reflect dynamics intrinsic to the proton's structure.
If the sea quarks were generated
solely by perturbative QCD evolution
via gluon splitting, the anti-quark distributions would be isospin
symmetric.  
Evidence for a difference between the  $\bar s(x)$ and $s(x)$ distributions
has also been claimed. \cite{Barone:1999yv}
Gluons carry a
significant fraction of the proton's spin as well as its momentum.  Since
gluon exchange between
valence quarks contributes to the
$p-\Delta$ mass splitting, it follows that the gluon distributions must be
cannot be solely accounted for by gluon bremsstrahlung from
individual quarks, the
process responsible for DGLAP evolutions of the structure functions.
Similarily,  in the case of
heavy quarks, $s\bar s$,
$c \bar c$, $b \bar b$, the diagrams in which the sea quarks are multiply
connected to
the valence quarks are intrinsic to the proton structure itself.  Thus
neither gluons nor sea quarks are solely
generated by DGLAP evolution, and one cannot define a resolution scale
$Q_0$ where the sea or
gluon degrees of freedom can be neglected.  There have also been
surprises associated with
the chirality distributions
$\Delta q = q_{\uparrow/\uparrow} - q_{\downarrow/\uparrow}$ of the valence
quarks  which again show that a simple valence quark
approximation to nucleon spin structure functions is far from the actual
dynamical situation.  For a recent discussion and references, see
Ref. \cite{Karliner:1999fn}.

A traditional focus of QCD has been on hard
inclusive processes and jet physics where perturbative methods and
leading-twist factorization provide predictions up to next-to-next-to
leading order (NNLO) with very good precision.  More recently, the
domain of reliable perturbative QCD predictions has been extended to
much more complex phenomena, such as a fundamental understanding of the hard
QCD BFKL pomeron in deep inelastic scattering at small $x_{bj}$ and hard
diffractive processes, such as $\gamma^* p \to \rho^0 p$.
In these lectures I will discuss applications of QCD where the
non-perturbative composition of hadrons in terms of their quark and
gluon degrees of freedom play a crucial role, for example the
$x_{bj}$-dependence of structure functions measured in deep inelastic
scattering, exclusive and semi-exclusive processes such as form factors,
two-photon processes, elastic scattering at fixed
$\theta_{cm}$, as well as the semi-leptonic decays of heavy hadrons.  The
analysis of QCD processes at the amplitude level is a challenging
problem, mixing issues involving non-perturbative and perturbative
dynamics.  However, a number of tools are available:

1. The Light-Cone Fock expansion provides a frame-independent
representation of a hadrons in terms of a set of wavefunctions
$\{\psi_{n/H}(x_i,\vec k_{\perp i},\lambda_i)\}$ describing its
composition into relativistic quark and gluon constituents.  The
light-cone wavefunctions can be derived from the eigensolutions of the
QCD Hamiltonian defined at fixed light-cone time $\tau = t + z/c.$
Structure functions are obtained from the sum over absolute squares of the
light-cone wavefunctions. Spacelike form factors and semi-leptonic decay
amplitudes can be written as exact identities in terms of the
convolution of the light-cone wavefunctions.

2. Factorization theorems for hard exclusive, semi-exclusive, and
diffractive processes allow a rigorous separation of soft
non-perturbative dynamics of the bound state hadrons from the hard
dynamics of the perturbatively-calculable quark-gluon scattering
amplitude $T_H^{(\Lambda)}$.  The key non-perturbative input is the
gauge and frame independent hadron distribution amplitude \cite{LB}
defined as the integral over transverse momenta of the valence (lowest
particle number) Fock wavefunction; \eg\ for the pion
\begin{equation}
\phi_\pi (x_i,Q) \equiv \int d^2k_\perp\, \psi^{(Q)}_{q\bar q/\pi}
(x_i, \vec k_{\perp i},\lambda)
\label{eq:f1}
\end{equation}
where the global cutoff $\Lambda$ is identified with the resolution $Q$.
The distribution amplitude controls leading-twist exclusive amplitudes
at high momentum transfer, and it can be related to the gauge-invariant
Bethe-Salpeter wavefunction at equal light-cone time $\tau = x^+$.  Thus
hard exclusive hadronic amplitudes such as quarkonium decay, heavy
hadron decay, and scattering amplitudes where the hadrons are scattered
with momentum transfer can be factorized as the convolution of the
light-cone Fock state wavefunctions with quark-gluon matrix elements
\cite{LB}
\begin{eqnarray}
\M_{\rm Hadron} &=& \prod_H \sum_n \int
\prod^{n}_{i=1} d^2k_\perp \prod^{n}_{i=1}dx_i \delta
\left(1-\sum^n_{i=1}x_i\right)\, \delta
\left(\sum^n_{i=1} \vec k_{\perp i}\right) \nonumber \\[2ex]
&& \times \psi^{(\Lambda)}_{n/H} (x_i,\vec k_{\perp i},\lambda_i)\,
T_H^{(\Lambda)} \ .
\label{eq:e1}
\end{eqnarray}
Here $T_H^{(\Lambda)}$ is the underlying quark-gluon subprocess
scattering amplitude, where the (incident or final) hadrons are replaced
by quarks and gluons with momenta $x_ip^+$, $x_i\vec p_{\perp}+\vec
k_{\perp i}$ and invariant mass above the separation scale $\M^2_n >
\Lambda^2$.

3. The logarithmic evolution of hadron distribution amplitudes
$\phi_H (x_i,Q)$ can be derived from the perturbatively-computable tail
of the valence light-cone wavefunction in the high transverse momentum
regime.\cite{LB} 

4. Conformal symmetry provides a template for
QCD predictions, leading to relations between observables which are
present even in a theory which is not scale invariant. Thus an
important guide in QCD analyses is to identify the
underlying conformal relations of QCD which are manifest if we drop quark
masses and effects due to the running of the QCD couplings. In fact, if
QCD has an infrared fixed point (vanishing of the Gell Mann-Low function
at low momenta), the theory will closely resemble a scale-free
conformally symmetric theory in many applications.

5.  Commensurate scale relations are perturbative QCD
predictions which relate observable to observable at fixed relative
scale, such as the ``generalized Crewther relation", which connects the
Bjorken and Gross-Llewellyn Smith deep inelastic scattering sum rules to
measurements of the $e^+ e^-$ annihilation cross section. The relations
have no renormalization scale or scheme ambiguity.  The coefficients in
the perturbative series for commensurate scale relations are identical to
those of conformal QCD; thus no infrared renormalons are present.   All
non-conformal effects are absorbed by fixing the ratio of the respective
momentum transfer and energy scales.  In the case of fixed-point theories,
commensurate scale relations relate both the ratio of couplings and the
ratio of scales as the fixed point is approached.

5. $\alpha_V$ Scheme.  A natural scheme for defining the QCD coupling in
exclusive and other processes is the $\alpha_V$ scheme defined from
heavy quark interactions.  All vacuum polarization corrections due to
fermion pairs are then automatically and analytically incorporated into
the Gell Mann-Low function, thus avoiding the problem of explicitly
computing and resumming quark mass corrections related to the running of
the coupling.

6. The Abelian Correspondence Principle.  One can consider QCD
predictions as analytic functions of the number of colors $N_C$ and
flavors $N_F$.  In particular, one can show at all orders of
perturbation theory that PQCD predictions reduce to those of an Abelian
theory at $N_C \to 0$ with ${\widehat \alpha} = C_F \alpha_s$ and
${\widehat N_F} = N_F/T C_F$ held fixed.\cite{Brodsky:1997jk}  There is
thus a deep connection between QCD processes and their corresponding QED
analogs.

\section{The Light-Cone Fock Expansion in QCD}

In a relativistic collision, an incident hadron projectile presents
itself as an ensemble of coherent states containing various numbers of
quark and gluon quanta.  Thus when a laser beam crosses a proton at
fixed ``light-cone" time $\tau = t+z/c= x^0 + x^z$, it encounters a
baryonic state with a given number of quarks, anti-quarks, and gluons in
flight with $n_q - n_{\bar q} = 3$.  The natural formalism for
describing these hadronic components in QCD is the light-cone Fock
representation obtained by quantizing the theory at fixed
$\tau$.\cite{PinskyPauli} For example, the proton state has the Fock
expansion
\begin{eqnarray}
\ket p &=& \sum_n \VEV{n\,|\,p}\, \ket n \nonumber \\
&=& \psi^{(\Lambda)}_{3q/p} (x_i,\vec k_{\perp i},\lambda_i)\,
\ket{uud} \\[1ex]
&&+ \psi^{(\Lambda)}_{3qg/p}(x_i,\vec k_{\perp i},\lambda_i)\,
\ket{uudg} + \cdots \nonumber
\label{eq:b1}
\end{eqnarray}
representing the expansion of the exact QCD eigenstate on a
non-interacting quark and gluon basis.  The probability amplitude for
each such $n$-particle state of on-mass shell quarks and gluons in a
hadron is given by a light-cone Fock state wavefunction
$\psi_{n/H}(x_i,\vec k_{\perp i},\lambda_i)$, where the constituents
have longitudinal light-cone momentum fractions
\begin{equation}
x_i = \frac{k^+_i}{p^+} = \frac{k^0+k^z_i}{p^0+p^z}\ ,\quad \sum^n_{i=1} x_i= 1
\ ,
\label{eq:c}
\end{equation}
relative transverse momentum
\begin{equation}
\vec k_{\perp i} \ , \quad \sum^n_{i=1}\vec k_{\perp i} = \vec 0_\perp \ ,
\label{eq:d}
\end{equation}
and helicities $\lambda_i.$  The effective lifetime of each configuration
in the laboratory frame is ${2 P_{lab}\over{\M}_n^2- M_p^2} $ where
\begin{equation}
\M^2_n = \sum^n_{i=1} \frac{k^2_\perp + m^2}{x} < \Lambda^2    \label{eq:a}
\end{equation}
is the off-shell invariant mass and $\Lambda$ is a global ultraviolet
regulator.  The form of $\psi^{(\Lambda)}_{n/H}(x_i, \vec k_{\perp
i},\lambda_i)$ is invariant under longitudinal boosts; \ie,\ the
light-cone wavefunctions expressed in the relative coordinates $x_i$ and
$k_{\perp i}$ are independent of the total momentum $P^+$, $\vec
P_\perp$ of the hadron.

The interactions of the proton reflects an average over the
interactions of its fluctuating states.  For example, a valence state
with small impact separation, and thus a small color dipole moment,
would be expected to interact weakly in a hadronic or nuclear target
reflecting its color transparency.  The nucleus thus filters
differentially different hadron
components.\cite{Bertsch,MillerFrankfurtStrikman} The ensemble
\{$\psi_{n/H}$\} of such light-cone Fock wavefunctions is a key concept
for hadronic physics, providing a conceptual basis for representing
physical hadrons (and also nuclei) in terms of their fundamental quark
and gluon degrees of freedom.  Given the $\psi^{(\Lambda)}_{n/H},$ we
can construct any spacelike electromagnetic or electroweak form factor
from the diagonal overlap of the LC wavefunctions.\cite{BD} Similarly,
the matrix elements of the currents that define quark and gluon
structure functions can be computed from the integrated squares of the
LC wavefunctions.\cite{LB,BrodskyLepage}
In general the LC ultraviolet regulators provide a
factorization scheme for elastic and inelastic scattering, separating
the hard dynamical contributions with invariant mass squared $\M^2 >
\Lambda^2$ from the soft physics with $\M^2 \le
\Lambda^2$ which is incorporated in the nonperturbative LC
wavefunctions.  (Similarly, the DGLAP evolution of quark and gluon
distributions can be derived by computing the variation of the Fock
expansion with respect to $\Lambda^2$.\cite{LB})

The light-cone Fock formalism is derived in the following way:  one
first constructs the light-cone time evolution operator $P^-=P^0-P^z$
and the invariant mass operator $H_{LC}= P^- P^+-P^2_\perp $ in
light-cone gauge $A^+=0$ from the QCD Lagrangian.  The total
longitudinal momentum $P^+ = P^0 + P^z$ and transverse momenta $\vec
P_\perp$ are conserved, \ie\ are independent of the interactions.  The
matrix elements of $H_{LC}$ on the complete orthonormal basis
$\{\ket{n}\}$ of the free theory $H^0_{LC} = H_{LC}(g=0)$ can then be
constructed.  The matrix elements $\VEV{n\,|\,H_{LC}\,|\,m}$ connect
Fock states differing by 0, 1, or 2 quark or gluon quanta, and they
include the instantaneous quark and gluon contributions imposed by
eliminating dependent degrees of freedom in light-cone gauge.

It is thus important to not only compute the spectrum of hadrons and
gluonic states, but also to determine the wavefunction of each QCD bound
state in terms of its fundamental quark and gluon degrees of freedom.
If we could obtain such nonperturbative solutions of QCD, then we could
compute the quark and gluon structure functions and distribution
amplitudes which control hard-scattering inclusive and exclusive
reactions as well as calculate the matrix elements of currents which
underlie electroweak form factors and the weak decay amplitudes of the
light and heavy hadrons.  The light-cone wavefunctions also determine
the multi-parton correlations which control the distribution of
particles in the proton fragmentation region as well as dynamical higher
twist effects.  Thus one can analyze not only the deep inelastic
structure functions but also the fragmentation of the spectator system.
Knowledge of hadron wavefunctions would also open a window to a deeper
understanding of the physics of QCD at the amplitude level, illuminating
exotic effects of the theory such as color transparency, intrinsic heavy
quark effects, hidden color, diffractive processes, and the QCD van der
Waals interactions.

Solving a quantum field theory such as QCD is clearly not easy.
However, highly nontrivial, one-space one-time relativistic quantum
field theories which mimic many of the features of QCD, have already
been completely solved using light-cone Hamiltonian
methods.\cite{PinskyPauli} Virtually any (1+1) quantum field theory can
be solved using the method of Discretized Light-Cone-Quantization
(DLCQ)  \cite{DLCQ,Brodsky:1991ir} where the matrix elements
$\VEV{n\,|\,H^{\Lambda)}_{LC}\,|\,m}$, are made discrete in momentum
space by imposing periodic or anti-periodic boundary conditions in
$x^-=x^0 - x^z$ and $\vec x_\perp$.  Upon diagonalization of $H_{LC}$,
the eigenvalues provide the invariant mass of the bound states and
eigenstates of the continuum.  In DLCQ, the Hamiltonian $H_{LC}$, which
can be constructed from the Lagrangian using light-cone time
quantization, is completely diagonalized, in analogy to Heisenberg's
solution of the eigenvalue problem in quantum mechanics.  The quantum
field theory problem is rendered discrete by imposing periodic or
anti-periodic boundary conditions.  The eigenvalues and eigensolutions
of collinear QCD then give the complete spectrum of hadrons, nuclei, and
gluonium and their respective light-cone wavefunctions.  A beautiful
example is ``collinear" QCD:  a variant of $QCD(3+1)$ defined by
dropping all of interaction terms in $H^{QCD}_{LC}$ involving transverse
momenta.\cite{Kleb} Even though this theory is effectively
two-dimensional, the transversely-polarized degrees of freedom of the
gluon field are retained as two scalar fields.  Antonuccio and Dalley
\cite{AD} have used DLCQ to solve this theory.  The diagonalization of
$H_{LC}$ provides not only the complete bound and continuum spectrum of
the collinear theory, including the gluonium states, but it also yields
the complete ensemble of light-cone Fock state wavefunctions needed to
construct quark and gluon structure functions for each bound state.
Although the collinear theory is a drastic approximation to physical
$QCD(3+1)$, the phenomenology of its DLCQ solutions demonstrate general
gauge theory features, such as the peaking of the wavefunctions at
minimal invariant mass, color coherence and the helicity retention of
leading partons in the polarized structure functions at $x\rightarrow
1$.  The solutions of the quantum field theory can be obtained for
arbitrary coupling strength, flavors, and colors.

In practice it is essential to introduce an ultraviolet regulator in
order to limit the total range of $\VEV{n\,|\,H_{LC}\,|\,m}$, such as
the ``global" cutoff in the invariant mass of the free Fock state.  One
can also introduce a ``local" cutoff to limit the change in invariant
mass $|\M^2_n-\M^2_m| < \Lambda^2_{\rm local}$ which provides
spectator-independent regularization of the sub-divergences associated
with mass and coupling renormalization.  Recently, Hiller, McCartor, and
I have shown\cite{Brodsky:1998hs} that the Pauli-Villars method has
advantages for regulating light-cone quantized Hamitonian theory.  We
show that Pauli-Villars fields satisfying three spectral conditions will
regulate the interactions in the ultraviolet, while at same time avoiding
spectator-dependent renormalization and preserving chiral symmetry.
Although gauge theories are usually quantized on the light-cone in
light-cone gauge $A^+=0$, it is also possible and interesting to quantize
the theory in Feynman gauge\cite{Srivastava:1999gi}.  Covariant gauges are
advantageous since they preserve the rotational symmetry of the gauge
interactions.

The natural renormalization scheme for the QCD coupling is
$\alpha_V(Q)$, the effective charge defined from the scattering of two
infinitely-heavy quark test charges.  
This is discussed in more detail below.
The renormalization scale can then
be determined from the virtuality of the exchanged momentum, as in the
BLM and commensurate scale
methods.\cite{BLM,CSR,BrodskyKataevGabaladzeLu,BJPR}  Similar
effective charges have been proposed by Watson\cite{Watson:1997fg} and
Czarnecki\etal \cite{Czarnecki:1998sz}

In principle, we could also construct the wavefunctions of QCD(3+1)
starting with collinear QCD(1+1) solutions by systematic perturbation
theory in $\Delta H$, where $\Delta H$ contains the terms which
produce particles at non-zero $k_\perp$, including the terms linear and
quadratic in the transverse momenta $\longvec k_{\perp i}$ which are
neglected in the Hamilton $H_0$ of collinear QCD.  We can write the exact
eigensolution of the full Hamiltonian as
\[ \psi_{(3+1)} = \psi_{(1+1)} + \frac{1}{M^2-H + i \epsilon }\,
\Delta H\, \psi_{(1+1)} \ , \]
where
\[\frac{1}{M^2-H + i \epsilon }
 = {\frac{1}{M^2-H_0 + i \epsilon }} +
{\frac{1}{M^2-H+ i \epsilon }}\Delta H{\frac{1}{M^2-H_0 + i \epsilon }} \]
can be represented as the continued iteration of the Lippmann Schwinger
resolvant.
Note that the matrix
$(M^2-H_0)^{-1}$ is known to any desired precision from the DLCQ solution
of collinear QCD.

\section{Electroweak Matrix Elements and Light-Cone Wavefunctions}

\vspace{.5cm}
\begin{figure}[htb]
\begin{center}
\leavevmode
\epsfbox{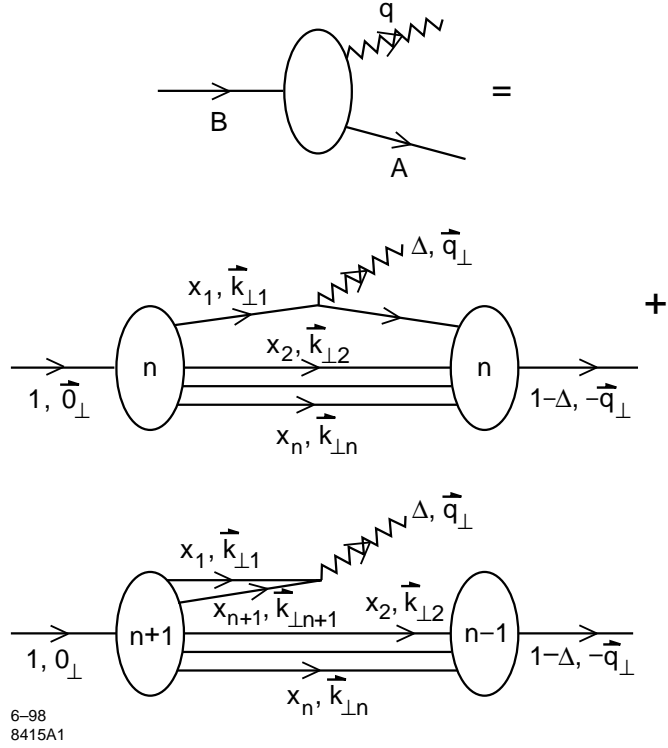}
\end{center}
\caption[*]{Exact representation of electroweak decays and time-like form
factors in the
light-cone Fock representation.
}
\label{fig1}
\end{figure}

Dae Sung Hwang and I have recently shown that exclusive semileptonic
$B$-decay amplitudes, such as $B\rightarrow A \ell \bar{\nu}$ can be
evaluated exactly in the light-cone formalism. \cite{Brodsky:1998hn}
These timelike decay matrix elements require the computation of the
diagonal matrix element $n \rightarrow n$ where parton number is
conserved, and the off-diagonal $n+1\rightarrow n-1$ convolution where
the current operator annihilates a $q{\bar{q'}}$ pair in the initial $B$
wavefunction.  See Fig.  \ref{fig1}.  This term is a consequence of the
fact that the time-like decay $q^2 = (p_\ell + p_{\bar{\nu}} )^2 > 0$
requires a positive light-cone momentum fraction $q^+ > 0$.  Conversely
for space-like currents, one can choose $q^+=0$, as in the
Drell-Yan-West representation of the space-like electromagnetic form
factors.  However, as can be seen from the explicit analysis of the form
factor in a perturbation model, the off-diagonal convolution can yield a
nonzero $q^+/q^+$ limiting form as $q^+ \rightarrow 0$.  This extra term
appears specifically in the case of ``bad" currents such as $J^-$ in
which the coupling to $q\bar q$ fluctuations in the light-cone
wavefunctions are favored.  In effect, the $q^+ \rightarrow 0$ limit
generates $\delta(x)$ contributions as residues of the $n+1\rightarrow
n-1$ contributions.  The necessity for such ``zero mode" $\delta(x)$ terms
has been noted by Chang, Root and Yan,\cite{CRY},Burkardt,\cite{BUR}
and Ji and Choi.\cite{Choi:1998nf}

The off-diagonal $n+1 \rightarrow n-1$ contributions give a new
perspective for the physics of $B$-decays.  A semileptonic decay
involves not only matrix elements where a quark changes flavor, but also
a contribution where the leptonic pair is created from the annihilation
of a $q {\bar{q'}}$ pair within the Fock states of the initial $B$
wavefunction.  The semileptonic decay thus can occur from the
annihilation of a nonvalence quark-antiquark pair in the initial hadron.
This feature will carry over to exclusive hadronic $B$-decays, such as
$B^0 \rightarrow \pi^-D^+$.  In this case the pion can be produced from
the coalescence of a $d\bar u$ pair emerging from the initial higher
particle number Fock wavefunction of the $B$.  The $D$ meson is then
formed from the remaining quarks after the internal exchange of a $W$
boson.

In principle, a precise evaluation of the hadronic matrix elements
needed for $B$-decays and other exclusive electroweak decay amplitudes
requires knowledge of all of the light-cone Fock wavefunctions of the
initial and final state hadrons.  In the case of model gauge theories
such as QCD(1+1)\cite{Horn} or collinear QCD \cite{AD} in one-space and
one-time dimensions, the complete evaluation of the light-cone
wavefunction is possible for each baryon or meson bound-state using the
DLCQ method.  It would be interesting to use such solutions as a model
for physical $B$-decays.

The existence of an exact formalism for electroweak matrix elements
gives a basis for systematic approximations and a control over
neglected terms.  For example, one can analyze exclusive semileptonic
$B$-decays  which  involve hard internal momentum transfer using a
perturbative QCD formalism patterned after the analysis of form factors at
large momentum transfer.\cite{LB}    The hard-scattering analysis proceeds by
writing each hadronic wavefunction as a sum of soft and hard contributions
\begin{equation}
\psi_n = \psi^{{\rm soft}}_n (\M^2_n < \Lambda^2) + \psi^{{\rm hard}}_n
(\M^2_n >\Lambda^2) ,
\end{equation}
where $\M^2_n $  is the invariant mass of the partons in the $n$-particle
Fock state and
$\Lambda$ is the separation scale.
The high internal momentum contributions to the wavefunction $\psi^{{\rm
hard}}_n $ can be calculated systematically from QCD perturbation theory
by iterating  the gluon exchange kernel.  The contributions from  high
momentum transfer exchange to the
$B$-decay amplitude can then be written as a convolution of a hard scattering
quark-gluon scattering amplitude $T_H$  with the distribution
amplitudes $\phi(x_i,\Lambda)$, the valence wavefunctions obtained by
integrating the
constituent momenta  up to the separation scale
${\cal M}_n < \Lambda < Q$.  This is the basis for the perturbative hard
scattering analyses.\cite{BHS,Sz,BALL,BABR}
In the exact analysis, one can
identify the hard PQCD contribution as well as the soft contribution from
the convolution of the light-cone wavefunctions. Furthermore, the hard
scattering contribution can be systematically improved.  For example, off-shell
effects can be retained in the evaluation of
$T_H$ by utilizing the exact light-cone energy denominators.

Given the solution
for the hadronic wavefunctions $\psi^{(\Lambda)}_n$ with $\M^2_n <
\Lambda^2$, one can construct the wavefunction in the hard regime with
$\M^2_n > \Lambda^2$ using projection operator techniques.\cite{LB} The
construction can be done perturbatively in QCD since only high invariant mass,
far off-shell matrix elements are involved.  One can use this method to
derive the physical properties of the LC wavefunctions and their matrix elements
at high invariant mass.  Since $\M^2_n = \sum^n_{i=1}
\left(\frac{k^2_\perp+m^2}{x}\right)_i $, this method also allows the derivation
of the asymptotic behavior of light-cone wavefunctions at large $k_\perp$, which
in turn leads to predictions for the fall-off of form factors and other
exclusive
matrix elements at large momentum transfer, such as the quark counting rules
for predicting the nominal power-law fall-off of two-body scattering amplitudes
at fixed
$\theta_{cm}.$\cite{BrodskyLepage} The phenomenological successes of these rules
can be understood within QCD if the coupling
$\alpha_V(Q)$ freezes in a range of relatively
small momentum transfer.\cite{BJPR}

\section{Other Applications of Light-Cone Quantization to QCD Phenomenology}

{\it Diffractive vector meson photoproduction.} The
light-cone Fock wavefunction representation of hadronic amplitudes
allows a simple eikonal analysis of diffractive high energy processes, such as
$\gamma^*(Q^2) p \to \rho p$, in terms of the virtual photon and the vector
meson Fock state light-cone wavefunctions convoluted with the $g p \to g p$
near-forward matrix element.\cite{BGMFS} One can easily show that only small
transverse size $b_\perp \sim 1/Q$ of the vector meson distribution
amplitude  is involved. The hadronic interactions are minimal, and thus the
$\gamma^*(Q^2) N \to
\rho N$ reaction can occur coherently throughout a nuclear target in reactions
without absorption or shadowing. The $\gamma^* A \to V A$ process thus
is a laboratory for testing QCD color transparency.\cite{BM}
This is discussed further in the next section.

{\it Regge behavior of structure functions.} The light-cone wavefunctions
$\psi_{n/H}$ of a hadron are not independent of each other, but rather are
coupled via the equations of motion. Antonuccio, Dalley and I
\cite{ABD} have used the constraint of finite ``mechanical'' kinetic energy to
derive ``ladder relations" which interrelate the light-cone wavefunctions of
states differing by one or two gluons.  We then use these relations to
derive the
Regge behavior of both the polarized and unpolarized structure functions at $x
\rightarrow 0$, extending Mueller's derivation of the BFKL hard
QCD pomeron from the properties of heavy quarkonium light-cone wavefunctions at
large $N_C$ QCD.\cite{Mueller}

{\it Structure functions at large $x_{bj}$.} The behavior of structure functions
where one quark has the entire momentum requires the knowledge of LC
wavefunctions
with $x \rightarrow 1$ for the struck quark and $x \rightarrow 0$ for the
spectators.  This is a highly off-shell configuration, and thus one can
rigorously derive quark-counting and helicity-retention rules for the
power-law behavior of
the polarized and unpolarized quark and gluon distributions in the $x
\rightarrow 1$ endpoint domain.  It is interesting to note that the evolution
of structure functions is minimal in this domain because the struck quark
is highly virtual as $x\rightarrow 1$; \ie\ the starting point $Q^2_0$ for
evolution
cannot be held fixed, but must be larger than a scale of order
$(m^2 + k^2_\perp)/(1-x)$.\cite{LB,BrodskyLepage,Dmuller}

{\it Intrinsic gluon and heavy quarks.}
The main features of the heavy sea quark-pair contributions of the Fock
state expansion of light hadrons can also be derived from perturbative QCD,
since $\M^2_n$ grows with
$m^2_Q$.  One identifies two contributions to the heavy quark sea, the
``extrinsic'' contributions which correspond to ordinary gluon splitting, and
the ``intrinsic" sea which is multi-connected via gluons to the valence quarks.
The intrinsic sea is thus sensitive to the hadronic bound state
structure.\cite{IC}  The maximal contribution of the
intrinsic heavy quark occurs at $x_Q \simeq {m_{\perp Q}/ \sum_i m_\perp}$
where $m_\perp = \sqrt{m^2+k^2_\perp}$;
\ie\ at large $x_Q$, since this minimizes the invariant mass $\M^2_n$.  The
measurements of the charm structure function by the EMC experiment are
consistent with intrinsic charm at large $x$ in the nucleon with a
probability of order $0.6 \pm 0.3 \% $.\cite{HSV} Similarly, one can
distinguish intrinsic
gluons which are associated with multi-quark interactions and extrinsic gluon
contributions associated with quark substructure.\cite{BS} One can also use this
framework to isolate the physics of the anomaly contribution to the Ellis-Jaffe
sum rule.

{\it Materialization of far-off-shell configurations.}
In a high energy hadronic collisions, the highly-virtual states of a hadron
can be
materialized into physical hadrons simply by the soft interaction of any of the
constituents.\cite{BHMT}  Thus a proton state with intrinsic charm $\ket{ u
u d \bar  c c}$ can be materialized, producing a $J/\psi$ at large $x_F$,
by the
interaction of a light-quark in the target.  The production
occurs on the front-surface of a target nucleus, implying an $A^{2/3}$
$J/\psi$ production cross section at large $x_F$ , which is consistent with
experiment, such as Fermilab experiments E772 and  E866.

{\it Rearrangement mechanism in heavy quarkonium decay.}
It is usually assumed that a heavy quarkonium state such as the
$J/\psi$ always decays to light hadrons via the annihilation of its heavy quark
constituents to gluons. However, as Karliner and I \cite{Brodsky:1997fj} have recently
shown, the transition  $J/\psi \to \rho
\pi$ can also occur by the rearrangement of the $c \bar c$ from the $J/\psi$
into the $\ket{ q \bar q c \bar c}$ intrinsic charm Fock state of the $\rho$ or
$\pi$. On the other hand, the overlap rearrangement integral in the
decay $\psi^\prime \to \rho \pi$ will be suppressed since the intrinsic
charm Fock state radial wavefunction of the light hadrons will evidently
not have nodes in its radial wavefunction. This observation gives a
natural explanation of the long-standing puzzle why the $J/\psi$ decays
prominently to two-body pseudoscalar-vector final states, whereas the
$\psi^\prime$ does not.

{\it Asymmetry of intrinsic heavy quark sea.}
The higher Fock state of the proton $\ket{u u d s \bar s}$ should
resemble  a $\ket{ K \Lambda}$ intermediate state, since this minimizes its
invariant mass $\M$.  In such a state, the
strange quark has a higher mean momentum fraction $x$ than the $\bar
s$. \cite{Warr,Signal,BMa} Similarly, the helicity intrinsic strange
quark in this configuration will be anti-aligned with the helicity of the
nucleon.%
\cite{Warr,BMa} This $Q \leftrightarrow \bar Q$ asymmetry
is a striking feature of the intrinsic heavy-quark sea.

{\it Comover phenomena.}
Light-cone wavefunctions describe not only the partons that interact in a hard
subprocess but also the associated partons freed from the projectile.  The
projectile partons which are comoving (\ie, which have similar rapidity) with
final state quarks and gluons can interact strongly producing (a) leading
particle effects, such as those seen in open charm hadroproduction; (b)
suppression of quarkonium\cite{BrodskyMueller} in favor of open heavy hadron
production, as seen in the E772 experiment; (c) changes in color configurations
and selection rules in quarkonium hadroproduction, as has been emphasized by
Hoyer and Peigne.\cite{Hoyer:1998ha}   All of these effects violate the
usual ideas of factorization for inclusive reactions. Further, more than
one parton from the
projectile can enter the hard subprocess, producing dynamical higher twist
contributions, as seen for example in
Drell-Yan experiments.\cite{BrodskyBerger,Brandenburg}

{\it Jet hadronization in light-cone QCD.}
One of the goals of nonperturbative analysis in QCD is to compute jet
hadronization from first principles.  The DLCQ solutions provide a possible
method to accomplish this.  By inverting the DLCQ solutions, we can write the
``bare'' quark state of the free theory as
$\ket{q_0} = \sum \ket n \VEV{n\,|\,q_0}$
 where now $\{\ket n\}$ are the exact DLCQ eigenstates of
$H_{LC}$, and
$\VEV{n\,|\,q_0}$ are the DLCQ projections of the eigensolutions.  The expansion
in automatically infrared and ultraviolet regulated if we impose global cutoffs
on the DLCQ basis:
$\lambda^2 < \Delta\M^2_n < \Lambda^2
$
where $\Delta\M^2_n = \M^2_n-(\Sigma \M_i)^2$.  It would be
interesting to study  jet hadronization at the amplitude level for
the existing DLCQ solutions to QCD (1+1) and collinear QCD.

{\it Hidden Color.}
The deuteron form factor at high $Q^2$ is sensitive to wavefunction
configurations where all six quarks overlap within an impact
separation $b_{\perp i} < {\cal O} (1/Q);$ the leading power-law
fall off predicted by QCD is $F_d(Q^2) = f(\alpha_s(Q^2))/(Q^2)^5$,
where, asymptotically, $f(\alpha_s(Q^2)) \propto
\alpha_s(Q^2)^{5+2\gamma}$.\cite{Brodsky:1976rz} The derivation of the
evolution equation for the deuteron distribution amplitude and its
leading anomalous dimension $\gamma$ is given in Ref. \cite{bjl83}
In general, the six-quark wavefunction of a deuteron
is a mixture of five different color-singlet states.  The dominant
color configuration at large distances corresponds to the usual
proton-neutron bound state.  However at small impact space
separation, all five Fock color-singlet components eventually
acquire equal weight, \ie, the deuteron wavefunction evolves to
80\%\ ``hidden color.'' The relatively large normalization of the
deuteron form factor observed at large $Q^2$ points to sizable
hidden color contributions.\cite{Farrar:1991qi}

{\it Spin-Spin Correlations in Nucleon-Nucleon
Scattering and the Charm~Threshold.}
One of the most striking anomalies in elastic proton-proton
scattering is the large spin correlation $A_{NN}$ observed at large
angles.\cite{krisch92} At $\sqrt s \simeq 5 $ GeV, the rate for
scattering with incident proton spins parallel and normal to the
scattering plane is four times larger than that for scattering with
anti-parallel polarization.  This strong polarization correlation can
be attributed to the onset of charm production in the intermediate
state at this energy.\cite{Brodsky:1988xw} The intermediate state $\vert u
u d u u d c \bar c \rangle$ has odd intrinsic parity and couples to
the $J=S=1$ initial state, thus strongly enhancing scattering when
the incident projectile and target protons have their spins parallel
and normal to the scattering plane.  The charm threshold can also
explain the anomalous change in color transparency observed at the
same energy in quasi-elastic $ p p$ scattering.  A crucial test is
the observation of open charm production near threshold with a
cross
section of order of $1 \mu$b.

\section{Hard Exclusive Reactions}

Exclusive hard-scattering reactions and hard diffractive reactions are now
giving
a valuable window into the structure and
dynamics of hadronic amplitudes.  Recent measurements of the
photon-to-pion transition form factor at CLEO,\cite{Gronberg:1998fj} the
diffractive dissociation of pions into jets at Fermilab,\cite{E791}
diffractive vector meson leptoproduction at Fermilab and HERA, and the new
program
of experiments on exclusive proton and deuteron processes at Jefferson
Laboratory
are now yielding fundamental information on hadronic wavefunctions,
particularly the
distribution amplitude of mesons.  Such information is also critical for
interpreting exclusive heavy hadron decays and the matrix elements and
amplitudes
entering $CP$-violating processes at the $B$ factories.

There has been much progress analyzing
exclusive and diffractive reactions at large momentum transfer from first
principles
in QCD.  Rigorous statements can be made on the basis of asymptotic freedom and
factorization theorems which separate the underlying hard quark and gluon
subprocess amplitude from the nonperturbative physics incorporated into the
process-independent hadron distribution amplitudes
$\phi_H(x_i,Q)$,\cite{LB} the valence light-cone
wavefunctions integrated over $k^2_\perp<Q^2$.  An important new
application is the
recent analysis of hard exclusive
$B$ decays by Beneke, {\it et al.}\cite{Beneke:1999br} Key features of such
analyses are: (a) evolution equations for distribution amplitudes which
incorporate the operator product expansion, renormalization group
invariance, and conformal symmetry;
\cite{LB,Brodsky:1980ny,Brodsky:1986ve,Muller:1994hg,Ball:1998ff,Braun:1999te}
(b) hadron helicity conservation which follows from the underlying chiral
structure of QCD;\cite{Brodsky:1981kj} (c) color transparency, which
eliminates corrections to hard exclusive amplitudes from initial and final state
interactions at leading power and reflects the underlying gauge theoretic basis
for the strong interactions;\cite{BM,Frankfurt:1992dx} and (d) hidden color
degrees of freedom in nuclear wavefunctions, which reflects the color
structure of hadron and nuclear wavefunctions.\cite{bjl83} There have also been
recent advances eliminating renormalization scale ambiguities in hard-scattering
amplitudes via commensurate scale
relations\cite{Brodsky:1995eh,Brodsky:1996tb,Brodsky:1999gm} which connect the
couplings entering exclusive amplitudes to the
$\alpha_V$ coupling which controls the QCD heavy quark
potential.\cite{Brodsky:1998dh} The postulate that the QCD coupling has an
infrared fixed-point can explain the applicability of
conformal scaling and dimensional counting
rules to physical QCD processes.\cite{BF,Matveev:1973ra,Brodsky:1998dh} The
field of analyzable exclusive processes has recently been expanded to a new
range
of QCD processes, such as electroweak decay
amplitudes, highly virtual diffractive processes such as
$\gamma^* p \to \rho p$,\cite{BGMFS,Collins:1997fb} and semi-exclusive
processes such as
$\gamma^* p \to \pi^+ X$ \cite{acw,Brodsky:1998sr,BB} where the $\pi^+$ is
produced in isolation at large $p_T$.

The natural
renormalization scheme
for the QCD coupling in hard exclusive processes is $\alpha_V(Q)$, the
effective charge defined from the scattering of two infinitely-heavy
quark test charges.  The renormalization scale can then be determined
from the virtuality of the exchanged momentum of the gluons, as in the BLM and
commensurate scale
methods.\cite{BLM,Brodsky:1995eh,Brodsky:1996tb,Brodsky:1999gm}

The main features of exclusive processes to leading power in the
transferred momenta are:

(1) The leading power fall-off is given by dimensional counting rules for
the hard-scattering amplitude: $T_H \sim 1/Q^{n-1}$, where $n$ is the total
number
of fields
(quarks, leptons, or gauge fields) participating in the hard
scattering.\cite{BF,Matveev:1973ra} Thus the reaction is dominated by
subprocesses
and Fock states involving the minimum number of interacting fields.  The
hadronic
amplitude follows this fall-off modulo logarithmic corrections from the
running of
the QCD coupling, and the evolution of the hadron distribution amplitudes.
In some
cases, such as large angle $p p \to p p $ scattering, pinch contributions from
multiple hard-scattering processes must also be
included.\cite{Landshoff:1974ew}
The general success of dimensional counting rules implies that the
effective coupling
$\alpha_V(Q^*)$ controlling the gluon exchange propagators in
$T_H$ are frozen in the infrared, \ie, have an infrared fixed point, since the
effective momentum transfers $Q^*$ exchanged by the gluons are often a
small fraction
of the overall momentum transfer.\cite{Brodsky:1998dh} The pinch contributions
are suppressed by a factor decreasing faster than a fixed power.\cite{BF}

(2) The leading power dependence is given by hard-scattering amplitudes $T_H$
which conserve quark helicity.\cite{Brodsky:1981kj,Chernyak:1999cj} Since the
convolution of $T_H$ with the light-cone wavefunctions projects out states with
$L_z=0$, the leading hadron amplitudes conserve hadron helicity; \ie, the
sum of
initial and final hadron helicities are conserved.

(3) Since the convolution of the hard scattering amplitude $T_H$ with the
light-cone
wavefunctions projects out the valence states with small impact parameter,
the essential part of the hadron wavefunction entering a hard exclusive
amplitude has
a small color dipole moment.  This leads to the absence of initial or final
state
interactions among the scattering hadrons as well as the color transparency.
of quasi-elastic interactions in a nuclear target.\cite{BM,Frankfurt:1992dx}
For example, the amplitude for diffractive vector meson photoproduction
$\gamma^*(Q^2) p \to \rho p$, can be written as convolution of the virtual
photon and
the vector meson Fock state light-cone wavefunctions the $g p \to g p$
near-forward matrix element.\cite{BGMFS} One can easily show that only
small transverse size $b_\perp \sim 1/Q$ of the vector meson distribution
amplitude is involved.  The sum over the interactions of the exchanged
gluons tend to
cancel reflecting its small color dipole moment.  Since the hadronic
interactions are
minimal,  the
$\gamma^*(Q^2) N \to
\rho N$ reaction at large $Q^2$ can occur coherently throughout a nuclear
target in
reactions without absorption or final state interactions.  The $\gamma^*A
\to V A$ process thus provides a natural framework for testing QCD color
transparency.  Evidence for color transparency in such reactions has been
found by Fermilab experiment E665.\cite{Adams:1997bh}

Diffractive multi-jet production in heavy
nuclei provides a novel way to measure the shape of the LC Fock
state wavefunctions and test color transparency.  For example, consider the
reaction
\cite{Bertsch,MillerFrankfurtStrikman,Frankfurt:1999tq}
$\pi A \rightarrow {\rm Jet}_1 + {\rm Jet}_2 + A^\prime$
at high energy where the nucleus $A^\prime$ is left intact in its ground
state.  The transverse momenta of the jets have to balance so that
$
\vec k_{\perp i} + \vec k_{\perp 2} = \vec q_\perp < {R^{-1}}_A \ ,
$
and the light-cone longitudinal momentum fractions have to add to
$x_1+x_2 \sim 1$ so that $\Delta p_L < R^{-1}_A$.  The process can
then occur coherently in the nucleus.  Because of color transparency,  \ie,
the cancelation of color interactions in a small-size color-singlet
hadron,  the valence wavefunction of the pion with small impact
separation will penetrate the nucleus with minimal interactions,
diffracting into jet pairs.\cite{Bertsch}
The $x_1=x$, $x_2=1-x$ dependence of
the di-jet distributions will thus reflect the shape of the pion distribution
amplitude; the $\vec k_{\perp 1}- \vec k_{\perp 2}$
relative transverse momenta of the jets also gives key information on
 the underlying shape of the valence pion
wavefunction.\cite{MillerFrankfurtStrikman,Frankfurt:1999tq} The QCD
analysis can be
confirmed by the observation that the diffractive nuclear amplitude
extrapolated to
$t = 0$ is linear in nuclear number $A$, as predicted by QCD color
transparency.  The integrated diffractive rate should scale as $A^2/R^2_A \sim
A^{4/3}$.  A diffractive dissociation experiment of this type, E791,  is now in
progress at Fermilab using 500 GeV incident pions on nuclear
targets.\cite{E791} The preliminary results from E791 appear to be consistent
with color transparency.  The momentum fraction distribution of the jets is
consistent with a valence light-cone wavefunction of the pion consistent with
the shape of the asymptotic distribution amplitude, $\phi^{\rm asympt}_\pi (x) =
\sqrt 3 f_\pi x(1-x)$.  As discussed below, data from
CLEO\cite{Gronberg:1998fj} for the
$\gamma
\gamma^* \rightarrow \pi^0$ transition form factor also favor a form for
the pion distribution amplitude close to the asymptotic solution\cite{LB}
to the perturbative QCD evolution
equation.\cite{Kroll,Rad,Brodsky:1998dh,Feldmann:1999wr,Schmedding:1999ap}
It will also be interesting to study diffractive tri-jet production using proton
beams
$ p A \rightarrow {\rm Jet}_1 + {\rm Jet}_2 + {\rm Jet}_3 + A^\prime $ to
determine the fundamental shape of the 3-quark structure of the valence
light-cone wavefunction of the nucleon at small transverse
separation.\cite{MillerFrankfurtStrikman} One interesting possibility is
that the distribution amplitude of the
$\Delta(1232)$ for $J_z = 1/2, 3/2$ is close to the asymptotic form $x_1
x_2 x_3$,  but that the proton distribution amplitude is more complex.
This would explain why the $p \to\Delta$ transition form factor appears to
fall faster at large $Q^2$ than the elastic $p \to p$ and the other $p \to
N^*$ transition form factors.\cite{Stoler:1999nj}
Conversely, one can use incident real and virtual photons:
$ \gamma^* A \rightarrow {\rm Jet}_1 + {\rm Jet}_2 + A^\prime $ to
confirm the shape of the calculable light-cone wavefunction for
transversely-polarized and longitudinally-polarized virtual photons.  Such
experiments will open up a direct window on the amplitude
structure of hadrons at short distances.

There are a large number of measured exclusive reactions in which the
empirical power
law fall-off predicted by dimensional counting and PQCD appears to be
accurate over a large range of momentum transfer.
These include processes such as the proton form factor, time-like meson pair
production in $e^+ e^-$ and $\gamma
\gamma$ annihilation, large-angle scattering processes such as pion
photoproduction
$\gamma p \to \pi^+ p$, and nuclear processes such as the deuteron form
factor at
large momentum transfer and deuteron photodisintegration.\cite{Brodsky:1976rz} A
spectacular example is the recent measurements at CESR of the photon to pion
transition form factor in the reaction $e \gamma \to e
\pi^0$.\cite{Gronberg:1998fj}
As predicted by leading twist QCD\cite{LB} $Q^2 F_{\gamma
\pi^0}(Q^2)$ is essentially constant for 1 GeV$^2 < Q^2 < 10$ GeV$^2.$
Further, the
normalization is consistent with QCD at NLO if one assumes that the pion
distribution
amplitude takes on the form $\phi^{\rm asympt}_\pi (x) =
\sqrt 3 f_\pi x(1-x)$ which is the asymptotic solution\cite{LB} to the
evolution equation for the pion
distribution amplitude.\cite{Kroll,Rad,Brodsky:1998dh,Schmedding:1999ap}

The measured deuteron form factor and the deuteron photodisintegration
cross section
appear to follow the leading-twist QCD predictions at large momentum
transfers in the
few GeV region.\cite{Holt:1990ze,Bochna:1998ca} The normalization of the
measured deuteron form factor is large compared to model calculations
\cite{Farrar:1991qi} assuming that the deuteron's six-quark wavefunction can be
represented at short distances with the color structure of two color
singlet baryons.
This provides indirect evidence for the presence of hidden color components as
required by PQCD.\cite{bjl83}

If the pion distribution amplitude is close to its asymptotic form, then one can
predict the normalization of exclusive amplitudes such as the spacelike
pion form factor $Q^2 F_\pi(Q^2)$.  Next-to-leading order
predictions are now becoming available which incorporate higher order
corrections
to the pion distribution amplitude as well as the hard scattering
amplitude.\cite{Muller:1994hg,Melic:1999hg,Szczepaniak:1998sa} However, the
normalization of the PQCD prediction for the pion form factor depends
directly on the
value of the effective coupling
$\alpha_V(Q^*)$ at momenta $Q^{*2} \simeq Q^2/20$.  Assuming
$\alpha_V(Q^*) \simeq 0.4$, the QCD LO prediction appears to be
smaller by approximately a factor of 2 compared to the presently available data
extracted
from the original pion electroproduction experiments from
CEA.\cite{Bebek:1976ww} A
definitive comparison will require a careful extrapolation to the pion pole and
extraction of the longitudinally polarized photon contribution of the $e p
\to \pi^+ n$ data.

An important debate has centered on whether processes such as the pion
and proton
form factors and elastic Compton scattering $\gamma p \to \gamma p$ might be
dominated by higher twist mechanisms until very large momentum
transfers.\cite{Isgur:1989iw,Radyushkin:1998rt,Bolz:1996sw} For example, if one
assumes that the light-cone wavefunction of the pion has the form
$\psi_{\rm soft}(x,k_\perp) = A \exp (-b {k_\perp^2\over x(1-x)})$, then the
Feynman endpoint contribution to the overlap integral at small $k_\perp$ and
$x \simeq 1$ will dominate the form factor compared to the hard-scattering
contribution until
very large $Q^2$.  However, the above form of $\psi_{\rm soft}(x,k_\perp)$
has no
suppression at $k_\perp =0$ for any $x$; \ie, the
wavefunction in the hadron rest frame does not fall-off at all for $k_\perp
= 0$ and
$k_z \to - \infty$.  Thus such wavefunctions do not represent well
soft QCD contributions.  Furthermore, endpoint contributions will be
suppressed
by the QCD Sudakov form factor, reflecting the fact that a near-on-shell
quark must
radiate
if it absorbs large momentum.  If the endpoint contribution dominates
proton Compton
scattering, then both photons will interact on the same
quark line in a local fashion and the
amplitude is real, in strong contrast to the QCD predictions which have a
complex
phase structure.  The perturbative QCD predictions\cite{Kronfeld:1991kp} for the
Compton amplitude phase can be tested in
virtual Compton scattering by interference with Bethe-Heitler
processes.\cite{Brodsky:1972vv} It should be
noted that there is no apparent endpoint contribution which
could explain the success of dimensional counting in large angle pion
photoproduction.

It is interesting to compare the corresponding calculations of form
factors of bound states in QED.  The soft wavefunction
is the Schr\"odinger-Coulomb solution $\psi_{1s}(\vec k) \propto (1 + {{\vec
p}^2/(\alpha m_{\rm red})^2})^{-2}$, and the full wavefunction,  which
incorporates transversely polarized photon exchange, only differs by
a factor $(1 + {\vec p}^2/m^2_{\rm red})$.  Thus the leading twist
dominance of form
factors in QED occurs at relativistic scales $Q^2 > {m^2_{\rm
red}} $.\cite{Brodsky:1989pv}
Furthermore, there are no extra relative factors of $\alpha$ in the
hard-scattering contribution.  If the QCD coupling $\alpha_V$ has an infrared
fixed point, then the fall-off of the valence wavefunctions of hadrons will have
analogous power-law
forms, consistent with the Abelian correspondence
principle.\cite{Brodsky:1997jk} If power-law wavefunctions are
indeed applicable to the soft domain of QCD then, the transition to
leading-twist
power law behavior will occur in the nominal hard perturbative QCD domain where
$Q^2 \gg \VEV{k^2_\perp}, m_q^2$.

\section{Semi-Exclusive Processes:  New Probes of Hadron Structure}

A new class of hard ``semi-exclusive''
processes of the form $A+B \to C + Y$, have been proposed as new probes of
QCD.\cite{BB,acw,Brodsky:1998sr} These processes are characterized
by a large momentum transfer $t= (p_A-p_C)^2$ and a large rapidity gap between
the final state particle $C$ and the inclusive system $Y$.
Here $A, B$
and $C$ can be hadrons or (real or virtual) photons.  The cross
sections for such processes factorize in terms of the distribution
amplitudes of $A$ and $C$ and the parton distributions in the target
$B$.  Because of this factorization semi-exclusive reactions provide a
novel array of generalized currents, which not only give insight into
the dynamics of hard scattering QCD processes, but also allow
experimental access to new combinations of the universal quark and
gluon distributions.

\begin{figure}[htb]
\begin{center}
  \leavevmode
  \epsfxsize=3.5in
 \epsfbox{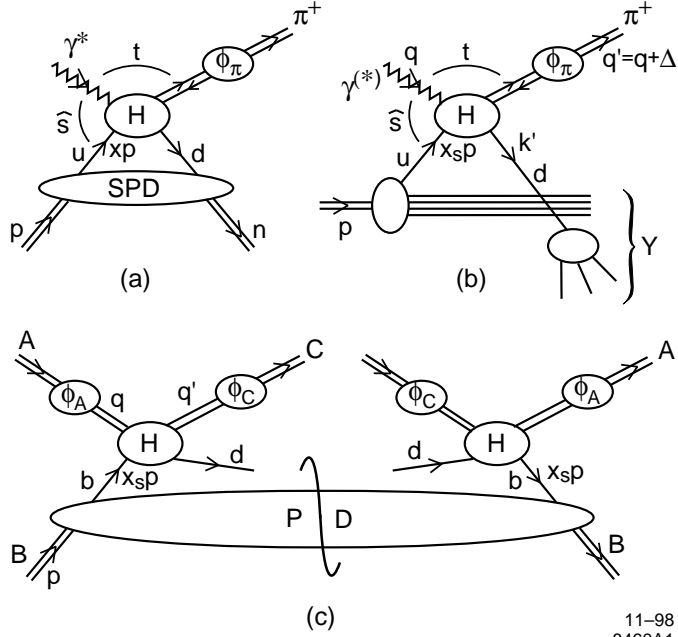}
\end{center}
\caption{{} (a): Factorization of $\gamma^* p \to \pi^+ n$ into a
 skewed parton distribution (SPD), a hard scattering $H$ and the pion
 distribution amplitude $\phi_\pi$.  (b): Semi-exclusive process
 $\gamma^{(*)} p \to \pi^+ Y$.  The $d$-quark produced in the hard
 scattering $H$ hadronizes independently of the spectator partons in
 the proton.  (c): Diagram for the cross section of a generic
 semi-exclusive process.  It involves a hard scattering $H$,
 distribution amplitudes $\phi_A$ and $\phi_C$ and a parton
 distribution (PD) in the target $B$.}
\label{one}
\end{figure}

QCD scattering amplitude for
deeply virtual exclusive processes like Compton scattering $\gamma^* p
\to \gamma p$ and meson production $\gamma^* p \to M p$ factorizes
into a hard subprocess and soft universal hadronic matrix
elements.  \cite{JiRad,Collins:1997fb,BGMFS}
For example, consider exclusive meson
electroproduction such as $e p \to e \pi^+ n$ (Fig.~\ref{one}a).  Here one
takes (as in DIS) the Bjorken limit of large photon virtuality, with
$x_B = Q^2/(2 m_p \nu)$ fixed, while the momentum transfer $t =
(p_p-p_n)^2$ remains small.  These processes involve `skewed' parton
distributions, which are generalizations of the usual parton
distributions measured in DIS.  The skewed distribution in Fig.~\ref{one}a
describes the emission of a $u$-quark from the proton target together
with the formation of the final neutron from the $d$-quark and the
proton remnants.  As the subenergy $\hat s$ of the scattering process
$\gamma^* u \to \pi^+ d$ is not fixed, the amplitude involves an
integral over the $u$-quark momentum fraction $x$.

An essential condition for the factorization of the deeply virtual
meson production amplitude of Fig.\ \ref{one}a is the existence of a large
rapidity gap between the produced meson and the neutron.  This
factorization remains valid if the
neutron is replaced with a hadronic system $Y$ of invariant mass $M_Y^2 \ll
W^2$, where $W$ is
the c.m.\ energy of the $\gamma^* p$ process.
For $M_Y^2 \gg m_p^2$ the momentum $k'$ of the $d$-quark in Fig.~\ref{one}b is
large with respect to the proton remnants, and hence it forms a jet.
This jet hadronizes independently of the other particles in the final
state if it is not in the direction of the meson, \ie, if the meson
has a large transverse momentum $q'_\perp = \Delta^{\phantom .}_\perp$
with respect to the photon direction in the $\gamma^* p$ c.m.  Then the
cross section for an inclusive system $Y$ can be calculated as in DIS,
by treating the $d$-quark as a final state particle.

The large $\dpt$ furthermore allows only transversally compact
configurations of the projectile $A$ to couple to the hard subprocess
$H$ of Fig.~\ref{one}b, as it does in exclusive processes.  \cite{LB} Hence the
above discussion applies not only to incoming virtual photons at large
$Q^2$, but also to real photons $(Q^2=0)$ and in fact to any hadron
projectile.

Let us then consider the general process $A+B\to C+Y$, where $B$ and
$C$ are hadrons or real photons, while the projectile $A$ can also be
a virtual photon.  In the semi-exclusive kinematic limit
$\lqcd^2,\, M_B^2,\, M_C^2 \ll
M_Y^2,\, \dpt^2 \ll W^2$
we have a large rapidity gap
$|y_C - y_d| = \log \frac{W^2}{\dpt^2 + M_Y^2}$
between $C$ and the parton $d$ produced in the hard scattering (see
Fig.~\ref{one}c). The cross section then factorizes into the form
\begin{eqnarray}
\lefteqn{ \frac{d\sigma}{dt\,dx_S}(A+B\to C+Y) }
 \hspace{4em} \nonumber \\
&=& \sum_{b} f_{b/B}(x_S,\mu^2) \frac{d\sigma}{dt} (A b \to C d)
 \eqcm \label{gencross}
\end{eqnarray}
where $t=(q-q')^2$ and $f_{b/B}(x_S,\mu^2)$ denotes the distribution
of quarks, antiquarks and gluons $b$ in the target $B$.  The momentum
fraction $x_S$ of the struck parton $b$ is fixed by kinematics to the
value
$x_S = \frac{-t}{M_Y^2-t}$
and the factorization scale $\mu^2$ is characteristic of the hard
subprocess $A b \to C d$.

It is conceptually helpful to regard the hard scattering amplitude $H$
in Fig.~\ref{one}c as a generalized current of momentum $q-q'=p_A - p_C$,
which interacts with the target parton $b$.  For $A= \gamma^*$ we
obtain a close analogy to standard DIS when particle $C$ is removed.
With $q' \to 0$ we thus find $-t \to Q^2$, $M_Y^2 \to W^2$, and see
that $x_S$ goes over into $x_B = Q^2 /(W^2 + Q^2)$.  The
possibility to control the value of $q'$ (and hence the momentum
fraction $x_S$ of the struck parton) as well as the quantum numbers of
particles $A$ and $C$ should make semi-exclusive processes a versatile
tool for studying hadron structure.  The cross section further depends
on the distribution amplitudes $\phi_A$, $\phi_C$ (\cf\ Fig.~\ref{one}c),
allowing new ways of measuring these quantities.

\section{ Conformal Symmetry as a Template}

Testing quantum chromodynamics to high precision is not easy.  Even in
processes involving high momentum transfer, perturbative QCD predictions
are complicated by questions of the convergence of the series,
particularly by the presence of ``renormalon" terms which grow as $n!$,
reflecting the uncertainty in the analytic form of the QCD coupling
at low scales.  Virtually all QCD processes are complicated by the
presence of dynamical higher twist effects, including power-law suppressed
contributions due to multi-parton correlations, intrinsic transverse
momentum, and finite quark masses.  Many of these effects are inherently
nonperturbative in nature and require knowledge of hadron wavefunction
themselves.  The problem of interpreting perturbative QCD predictions is
further compounded by theoretical ambiguities due to the apparent
freedom in the choice of renormalization schemes, renormalizations
scales, and factorization procedures.

A central principle of renormalization theory is that
predictions which relate physical observables to each other cannot depend
on theoretical conventions.  For example, one can use any renormalization
scheme, such as the modified minimal subtraction dimensional
regularization scheme, and any choice of renormalization scale $\mu$ to
compute perturbative series observables $A$ and $B$.  However, all traces
of the choices of the renormalization scheme and scale must disappear
when we algebraically eliminate the $\alpha_{\overline MS}(\mu)$
and directly relate $A$ to $B$.  This is the
principle underlying ``commensurate scale relations" (CSR)~\cite{CSR},
which are general leading-twist QCD predictions relating physical
observables to each other. For example, the ``generalized Crewther
relation",  which is discussed in more detail below,  provides a
scheme-independent relation between the QCD corrections to the Bjorken
(or Gross Llewellyn-Smith) sum rule for deep inelastic lepton-nucleon
scattering,  at a given momentum transfer $Q$, to the radiative
corrections to the annihilation cross section
$\sigma_{e^+ e^- \to
\rm hadrons}(s)$, at a corresponding ``commensurate" energy scale $\sqrt s$.
\cite{CSR,BGKL} The specific relation between the physical scales $Q$
and $\sqrt s$ reflects the fact
that the radiative corrections to each process have
distinct quark mass thresholds.

Any perturbatively calculable physical quantity can be used to define an
effective charge\cite{Grunberg,DharGupta,GuptaShirkovTarasov} by incorporating
the entire radiative correction into its definition.
For example, the $e^+ e^- \gamma^* \to {\rm
hadrons}$ annihilation to muon pair cross section ratio can be written
\begin{equation}
R_{e^+ e^-}(s) \equiv R^0_{e^+ e^-}(s) [ 1 +
{\alpha_R(s)\over \pi}] ,
\end{equation}
where $R^0_{e^+ e^-}$ is the prediction at Born level.
Similarly, we can define the entire radiative correction to the Bjorken
sum rule as the effective charge $\alpha_{g_1}(Q^2)$ where
$Q$ is the corresponding momentum transfer:
\begin{equation}
\int_0^1 d x \left[ g_1^{ep}(x,Q^2) - g_1^{en}(x,Q^2) \right]
   \equiv {1\over 6} \left|g_A \over g_V \right|
   C_{\rm Bj}(Q^2)
 = {1\over 6} \left|g_A
\over g_V \right| \left[ 1- {3\over 4} C_F{\alpha_{g_1}(Q^2) \over
   \pi} \right]  .
\end{equation}

By convention,
each effective charge is normalized to $\alpha_s$ in the weak coupling
limit.  One can define effective charges for virtually any quantity
calculable in perturbative QCD; \eg  moments of structure functions,
ratios of form factors, jet observables, and the effective potential
between massive quarks.  In the case of decay constants of the $Z$ or the
$\tau$, the mass of the decaying system serves as the physical scale in
the effective charge.  In the case of multi-scale observables, such as the
two-jet fraction in $e^+ e^-$ annihilation, the multiple arguments of the
effective coupling $\alpha_{2 jet}(s,y)$ correspond to the overall
available energy $s$ variables such as $y = \rm {max}_{ij}
(p_i+p_j)^2/s$ representing the maximum jet mass fraction.

Commensurate scale relations take the general form
\begin{equation}
\alpha_A(Q_A) = C_{AB}[\alpha_B(Q_B)] \ .
\end{equation}
The function $C_{AB}(\alpha_B)$ relates the observables $A$ and $B$ in
the conformal limit; \ie, $C_{AB}$ gives the functional dependence
between the effective charges which would be obtained if the theory had
zero $\beta$ function.  The conformal coefficients can be distinguished
from the terms associated with the $\beta$ function at each order in
perturbation theory from their color and flavor dependence, or by an
expansion about a fixed point.

The ratio
of commensurate scales is determined by the requirement that all terms
involving the $\beta$ function are incorporated into the arguments of the
running couplings, as in the original BLM procedure.
Physically, the ratio of scales corresponds to the fact that the physical
observables have different quark threshold and distinct sensitivities to
fermion loops.  More generally, the differing scales are in effect
relations between mean values of the physical scales which appear in loop
integrations.  Commensurate scale relations are transitive; \ie, given the
relation
between effective charges for observables $A$ and $C$ and $C$ and $B$,
the resulting between $A$ and $B$ is independent of $C$.  In particular,
transitivity implies $\Lambda_{AB} = \Lambda_{AC} \times \Lambda_{CB}$.

One can consider QCD predictions as functions of analytic
variables of the number of colors $N_C$ and flavors $N_F$.  For example,
one can show at all orders of perturbation theory that PQCD predictions
reduce to those of an Abelian theory at $N_C
\to 0$ with ${\widehat \alpha} = C_F \alpha_s$ and ${\widehat N_F} = N_f/T
C_F$ held fixed.  In particular, CSRs obey the ``Abelian correspondence
principle" in that they give the correct Abelian relations at $N_c \to 0.$

Similarly, commensurate scale relations  obey the ``conformal
correspondence principle":  the CSRs reduce to correct conformal
relations when $N_C$ and $N_F$ are tuned to produce zero
$\beta$ function. Thus conformal symmetry provides a {\it template} for
QCD predictions, providing relations between observables
which are present even in theories which are not scale invariant.  All
effects of the nonzero beta function are encoded in the appropriate
choice of relative scales $\Lambda_{AB} = Q_A/Q_B$.

The scale $Q$ which enters a given effective charge corresponds to a
physical momentum scale.  The total logarithmic derivative of each
effective charge effective charge $\alpha_A(Q)$ with respect to its
physical scale is given by the Gell Mann-Low equation:
\begin{equation}
{d\alpha_A(Q,m) \over d \log Q} = \Psi_A (\alpha_A(Q,m), Q/m) ,
\end{equation}
where the functional dependence of $\Psi_A$ is specific to its own
effective charge.  Here $m$ refers to the quark's pole mass.  The pole
mass is universal in that it does not depend on the choice of effective
charge.  The Gell Mann-Low relation is reflexive in that $\psi_A$ depends
on only on the coupling $\alpha_A$ at the same scale.  It should be
emphasized that the Gell Mann-Low equation deals with physical quantities
and is independent of the renormalization procedure and choice of
renormalization scale.  A central feature of quantum chromodynamics is
asymptotic freedom; \ie, the monotonic decrease of the QCD coupling
$\alpha_A(\mu^2)$ at large spacelike scales.  The empirical test of
asymptotic freedom is the verification of the negative sign of the Gell
Mann-Low function at large momentum transfer, which must be true for any
effective charge.

In perturbation theory,
\begin{equation}
\Psi_A = - \psi_A^{\{0\}} {\alpha_A^2\over \pi}
          - \psi_A^{\{1\}}{\alpha_A^3\over \pi^2}
          - \psi_A^{\{2\}} {\alpha_A^4\over \pi^3}   + \cdots
\end{equation}
At large scales $Q^2 \gg m^2$, the first two terms are universal and
identical to the first two terms of the $\beta$ function
$\psi_A^{\{0\}}=\beta_0 = {11 N_C \over 3} - {2 \over 3} N_F,
\psi_A^{\{1\}}=\beta_1,
$ whereas
$\psi_A^{\{n\}}$ for $n \ge 2$ is process dependent.  The quark mass
dependence of the $\psi$ function is analytic, and in the
case of $\alpha_V$ scheme is known to two loops.

The commensurate scale relation between $\alpha_A$ and $\alpha_B$ implies
an elegant relation between their conformal dependence $C_{AB}$ and their
respective Gell Mann Low functions:
\begin{equation}
\psi_B = {d C_{BA}\over d \alpha_A} \times \psi_A  .
\end{equation}
Thus given the result for $N_{F,V}(m/Q)$ in $\alpha_V$
scheme we can use the CSR to derive $N_{F,A}(m/Q)$ for any other effective
charge, at least to two loops.  The above relation also shows that if
one effective charge has a fixed point $\psi_A[\alpha_A(Q^{FP}_A)] = 0$,
then all effective charges $B$ have a corresponding fixed point
$\psi_B[\alpha_B(Q^{FP}_B)] = 0$ at the corresponding commensurate scale
and value of effective charge.

In quantum electrodynamics, the running coupling $\alpha_{QED}(Q^2)$,
defined from
the Coulomb scattering of two infinitely heavy test charges at the
momentum transfer
$t = -Q^2$, is
taken as the standard observable.  Is there a preferred effective charge
which we should use to characterize the
coupling strength in QCD?  In the case of QCD,  the heavy-quark potential
$V(Q^2)$ is defined via a Wilson loop from the interaction energy of
infinitely heavy quark and antiquark at momentum transfer $t = -Q^2.$ The
relation
$V(Q^2) = - 4 \pi C_F
\alpha_V(Q^2)/Q^2$ then defines
the effective charge $\alpha_V(Q).$
As in the corresponding case of Abelian QED, the scale $Q$ of the coupling
$\alpha_V(Q)$ is identified with the exchanged momentum.  Thus there is never
any ambiguity in the interpretation of the scale.  All vacuum polarization
corrections
due to fermion pairs are incorporated in $\alpha_V$ through the usual
vacuum polarization
kernels which depend on the physical mass thresholds.  Other
observables could be used to define the standard QCD coupling, such as the
effective
charge defined from heavy quark radiation.\cite{Uraltsev}

Commensurate
scale relations between $\alpha_V$ and the QCD
radiative corrections to other observables have no scale or scheme
ambiguity, even in
multiple-scale problems such as multi-jet production.  As is the case in
QED, the
momentum scale which appears as the argument of
$\alpha_V$ reflect the mean virtuality of the exchanged gluons.
Furthermore, we can
write a commensurate scale relation between $\alpha_V$ and an analytic
extension of
the
$\alpha_{\overline {MS}}$ coupling, thus transferring all of the
unambiguous scale-fixing
and analytic properties of the physical $\alpha_V$ scheme to the
$\overline {MS}$
coupling.

An elegant example is the relation between the rate for semi-leptonic
$B$-decay and $\alpha_V$:
\begin{equation}
\Gamma(b \to X_u \ell \nu) =
          {G_F^2 {\vert V_{ub}\vert}^2 M_b^2\over 192 \pi^3}
           \left[1-2.41 {\alpha_V(0.16 M_b)\over \pi} -
               1.43 {\alpha_V(0.16 M_b)\over \pi}^2 \right]  ,
\end{equation}
where $M_b$ is the scheme independent $b-$quark pole mass.  The
coefficient of $\alpha^2_{\overline MS}(\mu)$ in the usual expansion
with $\mu = m_b$ is 26.8.

Some other examples of CSR's at NLO:
\begin{equation}
\alpha_R(\sqrt s) =
\alpha_{g_1}(0.5 \sqrt s) -
{\alpha^2_{g1}(0.5 \sqrt s)\over \pi} +
{\alpha^3_{g1}(0.5 \sqrt s)\over \pi^2}
\end{equation}
\begin{equation}
\alpha_R(\sqrt s) =
\alpha_V(1.8 \sqrt s) + 2.08
{\alpha^2_V(1.8 \sqrt s)\over \pi} - 7.16
{\alpha^3_V(1.8 \sqrt s)\over \pi^2}
\end{equation}
\begin{equation}
\alpha_{\tau}(\sqrt s) =
\alpha_V(0.8 \sqrt s) + 2.08
{\alpha^2_V(0-.8 \sqrt s)\over \pi} - 7.16
{\alpha^3_V(0.8 \sqrt s)\over \pi^2}
\end{equation}
\begin{equation}
\alpha_{g1}(\sqrt s) =
\alpha_V(0.8 Q) + 1.08
{\alpha^2_V(0.8 Q)\over \pi} - 10.3
{\alpha^3_V(0.8 Q)\over \pi^2}
\end{equation}
For numerical purposes in each case we have used $N_F=5$ and $\alpha_V=
0.1$ to compute the NLO correction to the CSR scale.

Commensurate scale relations thus provide
fundamental and precise scheme-independent tests of QCD, predicting how
observables
track not only in relative normalization, but also in their commensurate scale
dependence.

\section{The Generalized Crewther Relation}

The generalized Crewther relation can be derived by calculating the
QCD radiative corrections to the deep inelastic sum rules and $R_{e^+ e^-}$ in a
convenient renormalization scheme such as the modified minimal subtraction
scheme $\overline{\rm MS}$.  One then algebraically eliminates
$\alpha_{\overline
{MS}}(\mu)$.
Finally, BLM scale-setting\cite{BLM} is used to eliminate the $\beta$-function
dependence of the coefficients.  The form of the resulting relation between the
observables thus matches the result which would have been obtained had QCD
been a
conformal theory with zero $\beta$ function.  The final result relating the
observables is independent of the choice of intermediate
$\overline{\rm MS}$ renormalization scheme.

More specifically, consider the Adler function\cite{Adler} for the $e^+
e^-$ annihilation cross section
\begin{equation} D(Q^2)=-12\pi^2 Q^2{d\over dQ^2}\Pi(Q^2),~
\Pi(Q^2) =-{Q^2\over 12\pi^2}\int_{4m_{\pi}^2}^{\infty}{R_{e^+ e^-}(s)ds\over
s(s+Q^2)}.
\end{equation}
The entire radiative correction to this function is defined as
the effective charge
$\alpha_D(Q^2)$:
\begin{eqnarray}
    D \left( Q^2/ \mu^2, \alpha_{\rm s}(\mu^2) \right) &=&
D \left (1, \alpha_{\rm s}(Q^2)\right) \label{3} \\
&\equiv&
    3 \sum_f Q_f^2 \left[ 1+ {3\over 4} C_F{\alpha_D(Q^2) \over \pi}
                   \right]
    +( \sum_f Q_f)^2C_{\rm L}(Q^2) \nonumber \\
&\equiv& 3 \sum_f Q_f^2 C_D(Q^2)+( \sum_f Q_f)^2C_{\rm L}(Q^2),
\nonumber
\end{eqnarray}
where $C_F={N_C^2-1\over 2 N_C}. $
The coefficient $C_{\rm L}(Q^2)$ appears at the third order in
perturbation theory and is related to the ``light-by-light scattering type"
diagrams.  (Hereafter $\alpha_{\rm s}$ will denote the ${\overline{\rm MS}}$
scheme strong coupling constant.)

It is straightforward to algebraically relate $\alpha_{g_1}(Q^2)$ to
$\alpha_D(Q^2)$ using the known expressions to three loops in the
$\overline{\rm MS}$ scheme.  If one chooses the renormalization scale to resum
all of the quark and gluon vacuum polarization corrections into
$\alpha_D(Q^2)$,  then
the final result turns out to be remarkably simple\cite{BGKL}
 $(\widehat\alpha = 3/4\, C_F\ \alpha/\pi):$
\begin{equation}
\widehat{\alpha}_{g_1}(Q)=\widehat{\alpha}_D( Q^*)-
\widehat{\alpha}_D^2( Q^*)+\widehat{\alpha}_D^3( Q^*) +
\cdots,
\end{equation}
where
\begin{eqnarray}
\ln \left({ {Q}^{*2} \over Q^2} \right) &=&
{7\over 2}-4\zeta(3)+\left(\frac{\alpha_D ( Q^*)}{4\pi}
\right)\Biggl[ \left(
            {11\over 12}+{56\over 3} \zeta(3)-16{\zeta^2(3)}
      \right) \beta_0\cr &&
      +{26\over 9}C_{\rm A}
      -{8\over 3}C_{\rm A}\zeta(3)
      -{145\over 18} C_{\rm F}
      -{184\over 3}C_{\rm F}\zeta(3)
      +80C_{\rm F}\zeta(5)
\Biggr].
\label{EqLogScaleRatio}
\end{eqnarray}
where in QCD, $C_{\rm A}=N_C = 3$ and $C_{\rm F}=4/3$.
This relation shows how
the coefficient functions for these two different processes are
related to each other at their respective commensurate scales.  We emphasize
that the $\overline{\rm MS}$ renormalization
scheme is used only for calculational convenience; it serves simply as an
intermediary between observables.  The renormalization group
ensures
that the forms of the CSR relations in perturbative QCD are independent
of the choice of an intermediate renormalization scheme.

The Crewther relation was originally derived assuming that the theory is
conformally
invariant; \ie, for zero $\beta$ function.  In the physical case, where the
QCD coupling runs,  all non-conformal effects are resummed into the
energy and momentum transfer scales of
the effective couplings $\alpha_R$ and $\alpha_{g1}$.  The general
relation between these two effective charges for non-conformal
theory thus takes the form of a geometric series
\begin{equation}
        1- \widehat \alpha_{g_1} =
\left[ 1+ \widehat \alpha_D( Q^*)\right]^{-1} \ .
\end{equation}
We have dropped the small light-by-light scattering contributions.
This is again a special advantage of relating observable to
observable.
The coefficients are independent of
color and are the same in Abelian, non-Abelian, and conformal gauge theory.  The
non-Abelian structure of the theory is reflected in the expression for the scale
${Q}^{*}$.

Is experiment consistent with the generalized Crewther relation?  Fits
\cite{MattinglyStevenson} to the experimental measurements of the
$R$-ratio above the thresholds for the production of $c\overline{c}$ bound
states
provide the empirical constraint:
$\alpha_{R}({\sqrt s} =5.0~{\rm GeV})/\pi \simeq 0.08\pm 0.03.$
The prediction for the effective coupling for
the deep inelastic sum rules at the commensurate momentum transfer $Q$
is then
$\alpha_{g_1}(Q=12.33\pm 1.20~{\rm GeV})/\pi
\simeq \alpha_{\rm GLS}(Q=12.33\pm 1.20~{\rm GeV})/\pi
\simeq 0.074\pm 0.026.$
Measurements of the Gross-Llewellyn Smith sum rule have so far only been
carried out
at relatively small values of $Q^2$\cite{CCFRL1,CCFRL2};
however, one can use the results of the theoretical
extrapolation\cite{KS} of the experimental data presented in\cite{CCFRQ}:
$ \alpha_{\rm GLS}^{\rm extrapol}(Q=12.25~{\rm GeV})/\pi\simeq 0.093\pm 0.042.$
This range overlaps with the prediction from the generalized
Crewther relation.  It is clearly important to have higher precision
measurements to
fully test this fundamental QCD prediction.

\section{General Form of Commensurate Scale Relations}

In general, commensurate scale
relations connecting the effective charges for observables $A$ and $B$ have the
form
\begin{equation}
\alpha_A(Q_A) =
\alpha_B(Q_B) \left(1 + r^{(1)}_{A/B} {\alpha_B(Q_B)\over \pi} + r^{(2)}_{A/B}
{\alpha_B(Q_B)\over
\pi}^2 + \cdots\right),
\label{eq:CSRg}
\end{equation}
where the coefficients $r^{{n}}_{A/B}$
are
identical to the coefficients obtained in a con\-formally invariant theory
with $\beta_B(\alpha_B) \equiv (d/d\ln Q^2) \alpha_B(Q^2) = 0$.
The ratio of the scales $Q_A/Q_B$ is thus fixed by the requirement that
the couplings sum all of the effects of the non-zero $\beta$ function.  In
practice the
NLO and NNLO coefficients and relative scales can be identified from the flavor
dependence of the perturbative series; \ie\ by shifting scales such that the
$N_F$-dependence associated with $\beta_0 = 11/3 C_A - 4/3 T_F N_F$ and
$\beta_1 =
-34/3 C_A^2 + {20\over 3} C_A T_F N_F + 4 C_F T_F N_F$
does not appear in
the coefficients.  Here $C_A=N_C$, $C_F=(N^2_C-1)/2N_C$ and $T_F=1/2$.
The shift in scales which gives conformal coefficients in effect pre-sums
the large and strongly divergent terms in the PQCD series which grow as
$n!  (\beta_0
\alpha_s)^n$, \ie, the infrared renormalons associated with coupling-constant
renormalization.\cite{tHooft,Mueller,LuOneDim,BenekeBraun}

The
renormalization scales $Q^*$ in the BLM method are physical in the sense that
they reflect the mean virtuality of the gluon propagators.  This
scale-fixing procedure is consistent with scale fixing in QED, in agreement
with
in the Abelian limit, $N_C \to
0$.\cite{BLM,Brodsky:1997jk,LepageMackenzie,Neubert,BallBenekeBraun}
The ratio of scales
$\lambda_{A/B} = Q_A/Q_B$ guarantees that the
observables $A$ and $B$ pass through new quark thresholds at the same physical
scale.  One can also show that the commensurate scales satisfy the
transitivity rule
$\lambda_{A/B} = \lambda_{A/C} \lambda_{C/B},$ which ensures that predictions
are independent of the choice of an intermediate renormalization scheme or
intermediate observable $C.$

In general, we can write the relation between any two effective charges at
arbitrary
scales
$\mu_A$ and
$\mu_B$ as a correction to the corresponding relation obtained in a
conformally invariant
theory:
\begin{equation}
\alpha_A(\mu_A) = C_{AB}[\alpha_B(\mu_B)] +
\beta_B[\alpha_B(\mu_B)] F_{AB}[\alpha_B(\mu_B)]
\label{eq:ak}
\end{equation}
where
\begin{equation}
 C_{AB}[\alpha_B] = \alpha_B + \sum_{n=1} C_{AB}^{(n)}\alpha^n_B
\label{eq:al}
\end{equation}
is the functional relation when $\beta_B[\alpha_B]=0$.  In fact, if $\alpha_B$
approaches a fixed point $\bar\alpha_B$ where
$\beta_B[\bar\alpha_B]=0$,
then $\alpha_A$ tends to a fixed point given by
\begin{equation}
\alpha_A \to \bar\alpha_A = C_{AB}[\bar\alpha_B].
\label{eq:am}
\end{equation}
The commensurate scale relation for observables $A$ and $B$ has a similar
form, but
in this case the relative scales are fixed such that the non-conformal term
$F_{AB}$ is
zero.
Thus the commensurate scale relation $\alpha_A(Q_A) = C_{AB}[\alpha_B(Q_B)]$
at general commensurate scales is also the relation connecting the values
of the fixed
points for any two effective charges or schemes.  Furthermore, as
$\beta\rightarrow 0$,
the ratio of commensurate scales $Q^2_A/Q^2_B$ becomes the ratio of
fixed point scales $\bar Q^2_A/\bar Q^2_B$ as one approaches
the fixed point regime.

\section{Implementation of $\alpha_V$ Scheme}
\unboldmath

The effective charge $\alpha_V(Q)$ provides a physically-based alternative to
the usual modified minimal
subtraction ($\overline{\mbox{MS}}$) scheme.   All vacuum polarization
corrections due to fermion pairs are incorporated in $\alpha_V$ through
the usual vacuum polarization kernels which depend on the physical mass
thresholds.  When continued to time-like momenta, the coupling has the
correct analytic dependence dictated by the production thresholds in the
crossed channel.  Since $\alpha_V$ incorporates quark mass effects
exactly, it avoids the problem of explicitly computing and resumming
quark mass corrections which are related to the running of the coupling.
Thus the effective number of flavors $N_F(Q/m)$ is an analytic function
of the scale $Q$ and the quark masses $m$.  The effects of finite quark
mass corrections on the running of the strong coupling were first
considered by De R{\' u}jula and Georgi~\cite{derujula}
within the momentum subtraction schemes (MOM)
(see also Georgi and Politzer~\cite{Georgi_Politzer},
Shirkov and collaborators~\cite{shirkov}, and Ch{\'y}la~\cite{chyla}).

One important advantage of the physical charge approach is
its inherent gauge invariance to all orders in perturbation theory. This
feature is not manifest in massive $\beta$-functions defined in non-physical
schemes such as the MOM schemes.  A second, more practical,
advantage is the automatic decoupling of heavy quarks according to the
Appelquist-Carazzone theorem\cite{ac}.

By employing the commensurate scale relations
other physical observables
can be expressed in terms of the analytic
coupling $\alpha_V$ without scale or scheme ambiguity.
This way the quark mass threshold effects
in the running of the coupling are taken into account by
 utilizing the mass dependence of the physical
$\alpha_V$ scheme.  In effect, quark thresholds are treated analytically to all
orders in $m^2/Q^2$; {\it i.e.},  the evolution of the physical $\alpha_V$
coupling in the intermediate regions reflects the actual mass dependence of a
physical effective charge and the analytic properties of particle production.
 Furthermore, the definiteness of the
dependence in the quark masses automatically constrains the
scale $Q$ in the argument of the coupling.
There is thus no scale ambiguity in perturbative expansions in $\alpha_V$.

In the conventional $\overline{\mbox{MS}}$ scheme,
the coupling is independent of the
quark masses since the quarks are treated as either massless or
infinitely heavy with respect to the running of the coupling.  Thus one
formulates different effective theories depending on the effective number of
quarks which is governed by the scale $Q$; the massless $\beta$-function
is used to describe the running in between the flavor thresholds.
These different theories are then matched to each other by imposing matching
conditions at the scale where the effective number of flavors is changed
(normally the quark masses).  The dependence on the matching scale
 can be made arbitrarily small by calculating the matching conditions
to high enough order.  For physical observables one can then include the effects
of finite quark masses by making a higher-twist expansion in $m^2/Q^2$ and
$Q^2/m^2$ for light and heavy quarks, respectively.  These higher-twist
contributions have to be calculated for each observables separately, so
that in principle one requires an all-order resummation of the
mass corrections to the effective Lagrangian to give correct
results.

The specification of the coupling and renormalization scheme also depends
on the definition of the quark mass.  In contrast to QED where the
on-shell mass provides a natural definition of lepton masses,  an
on-shell definition for quark masses is complicated by the confinement
property of QCD.  In this paper we will use the pole mass $m(p^2=m^2)=m$
which has the advantage of being scheme and renormalization-scale invariant.
Furthermore, when combined with the $\alpha_V$ scheme,
the pole mass gives predictions which are
free of the leading renormalon ambiguity.

A technical complication of massive schemes is that one cannot easily obtain
analytic solutions of renormalization group equations to the massive $\beta$
function, and the Gell-Mann Low function is scheme-dependent even at lowest
order.

In a recent paper we have presented a two-loop analytic extension of the
$\alpha_V$-scheme based on the recent results of Ref.~\cite{melles98}.
The mass effects are in principle treated
exactly to two-loop order and are only limited in practice by
the uncertainties from numerical integration.
The desired features of gauge invariance and decoupling are manifest in
the form of the two-loop Gell-Mann Low function, and we give a simple
fitting-function which interpolates smoothly the exact two-loop results
obtained by using
the adoptive Monte Carlo integrator VEGAS\cite{vegas}.  Strong
consistency checks of the results are performed by comparing the Abelian
limit to the well known QED results in the on-shell scheme.  In addition,
the massless as well as the decoupling limit are reproduced exactly, and
the two-loop Gell-Mann Low function is shown to be renormalization scale
independent.

\begin{figure}[htb]
\begin{center}
\leavevmode
\epsfxsize=5in
\epsfbox{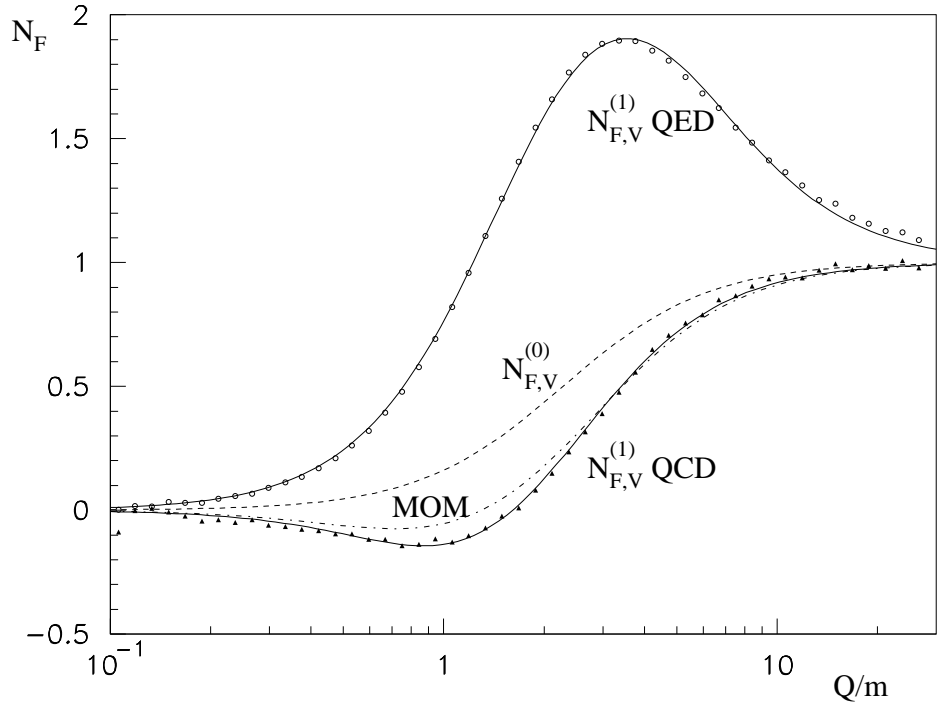}
\end{center}
\caption{The numerical results for the gauge-invariant $N_{F,V}^{(1)}$ in QED
(open circles) and QCD (triangles) with the best $\chi^2$ fits 
superimposed respectively.  The dashed
line shows the one-loop $N_{F,V}^{(0)}$ function .
For comparison we
also show the gauge dependent two-loop result obtained in MOM schemes
(dash-dot) \protect\cite{yh,jt}.  At large $\frac{Q}{m}$ the theory becomes
effectively massless, and both schemes agree as expected.  The figure also
illustrates the decoupling of heavy quarks at small $\frac{Q}{m}$.}
\label{fig:nfV}
\end{figure}

The results of our numerical calculation of $N_{F,V}^{(1)}$ in the
$V$-scheme for QCD and QED are shown in Fig.~\ref{fig:nfV}.
The decoupling of heavy quarks becomes manifest at small $Q/m$, and
the massless limit is attained for large $Q/m$.  The QCD form actually
becomes negative at moderate values of $Q/m$, a novel feature of
the anti-screening non-Abelian contributions.  This property is
also present in the (gauge dependent) MOM results.
In contrast, in Abelian QED the two-loop contribution to
the effective number of flavors becomes larger than 1 at intermediate
values of $Q/m$.  We also display the one-loop
contribution $N^{(0)}_{F,V} \left( \frac{Q}{m} \right)$ which
monotonically interpolates between the decoupling and massless limits.
The solid curves displayed in
Fig.~\ref{fig:nfV} shows that the parameterizations 
which we used for fitting the numerical results are quite accurate.

The relation of $\alpha_V(Q^2)$ to the conventional $\overline {MS}$
coupling is now known to NNLO,\cite{Peter}
but for clarity in this section only the NLO relation will be used.  The
commensurate scale relation is given by\cite{bgmr}
\begin{eqnarray}
\label{eq:csrmsovf}
\alpha_{\overline{\mbox{\tiny MS}}}(Q)
& = & \alpha_V(Q^{*}) + \frac{2}{3}N_C{\alpha_V^2(Q^{*}) \over \pi}
\nonumber \\
& = & \alpha_V(Q^{*}) + 2{\alpha_V^2(Q^{*}) \over \pi}\  ,
\end{eqnarray}
which is valid for $Q^2 \gg m^2$.  The coefficients in the
perturbation expansion have their conformal values, \ie, the same coefficients
would occur even if the theory had been conformally invariant with $\beta=0$.
The commensurate scale is given by
\begin{eqnarray}
Q^* & = & Q\exp\left[\frac{5}{6}\right] \ .
\end{eqnarray}
 The scale in the $\overline {MS}$ scheme is thus a factor
$\sim 0.4$ smaller than the physical scale.  The coefficient $2 N_C/3$ in
the NLO
coefficient is a feature of the non-Abelian couplings of QCD; the same
coefficient occurs even if the theory were conformally invariant with
$\beta_0=0.$

Using the above QCD results, we can transform any NLO prediction
given in $\overline{MS}$ scheme to a scale-fixed expansion in
$\alpha_V(Q)$.
We can also derive the connection between the $\overline{MS}$ and $\alpha_V$
schemes for Abelian perturbation theory using the limit $N_C \to 0$ with
$C_F\alpha_s$ and $N_F/C_F$ held fixed.\cite{Brodsky:1997jk}

The use of $\alpha_V$ and related physically defined effective charges such as
$\alpha_p$ (to NLO the effective charge defined from the (1,1) plaquette,
$\alpha_p$ is the same as $\alpha_V$) as expansion parameters has been
found to be valuable in lattice
gauge theory, greatly increasing the convergence of perturbative expansions
relative to
those using the bare lattice coupling.\cite{LepageMackenzie} Recent lattice
calculations of the
$\Upsilon$- spectrum\cite{Davies} have been used with BLM
scale-fixing to determine a NLO normalization
of the static heavy quark potential:  $
\alpha_V^{(3)}(8.2 \GeV) = 0.196(3)$ where the effective number of light
flavors is
$n_f = 3$.  The
corresponding modified minimal subtraction coupling evolved to the
$Z$ mass and five flavors is $ \alpha_{\overline{MS}}^{(5)}(M_Z) =
0.1174(24)$.  Thus a high precision value for $\alpha_V(Q^2)$ at a specific
scale is
available from lattice gauge theory.  Predictions for other QCD observables
can be
directly referenced to this value without the scale or scheme ambiguities,
thus greatly
increasing the precision of QCD tests.

One can also use $\alpha_V$ to characterize the coupling which appears in
the hard
scattering contributions of exclusive process amplitudes at large momentum
transfer,
such as elastic hadronic form factors, the photon-to-pion transition form
factor at
large momentum transfer\cite{BLM,BJPR} and exclusive weak decays of heavy
hadrons.\cite{Henley} Each gluon
propagator with four-momentum $k^\mu$ in the hard-scattering quark-gluon
scattering amplitude $T_H$ can be associated with the coupling
$\alpha_V(k^2)$ since the
gluon exchange propagators closely resembles the interactions encoded in the
effective potential $V(Q^2)$.  [In Abelian theory this is exact.]
Commensurate scale
relations can then be
established which connect the hard-scattering subprocess amplitudes which
control exclusive processes to other QCD observables.

We can anticipate that eventually
nonperturbative
methods such as lattice gauge theory or discretized light-cone quantization will
provide a complete form for the heavy quark potential in $QCD$.  It
is reasonable to assume that $\alpha_V(Q)$ will not diverge at small space-like
momenta.  One possibility is that $\alpha_V$ stays relatively constant
$\alpha_V(Q) \simeq 0.4$ at low momenta, consistent with fixed-point behavior.
There is, in fact, empirical evidence for freezing of the $\alpha_V$ coupling
from the observed systematic dimensional scaling behavior of exclusive
reactions.\cite{BJPR} If this is in fact the case, then the range of QCD
predictions can be extended to quite low momentum scales, a regime normally
avoided because of the apparent singular structure of perturbative
extrapolations.

There are a number of other advantages of the $V$-scheme:
\begin{enumerate}
\item
Perturbative expansions in $\alpha_V$ with the scale set by the momentum
transfer cannot
have any
$\beta$-function dependence in their coefficients since all running
coupling effects are
already summed into the definition of the potential.  Since
coefficients involving
$\beta_0$ cannot occur in an expansions in $\alpha_V$,  the divergent infrared
renormalon series of the form $\alpha^n_V\beta_0^n n!$ cannot occur.  The
general convergence properties of the scale $Q^*$ as an expansion in $\alpha_V$
is not known.\cite{Mueller}

\item
The effective coupling $\alpha_V(Q^2)$ incorporates vacuum polarization
contributions with finite fermion masses.  When continued to time-like
momenta, the coupling has the correct analytic dependence dictated by the
production thresholds in the $t$ channel.  Since $\alpha_V$ incorporates quark
mass effects exactly, it avoids the problem of explicitly computing and
resumming
quark mass corrections.

\item
The $\alpha_V$ coupling is the natural expansion parameter for
processes involving non-relativistic momenta, such as heavy quark production at
threshold where the Coulomb interactions, which are enhanced at low relative
velocity $v$ as $\pi \alpha_V/v$, need to be
re-summed.\cite{Voloshin,Hoang,Fadin}
The effective Hamiltonian for nonrelativistic QCD is thus most naturally
written in
$\alpha_V$ scheme.
The threshold corrections
to heavy quark production in $e^+ e^-$ annihilation depend on
$\alpha_V$ at specific scales $Q^*$.  Two distinct ranges of scales arise
as arguments of
$\alpha_V$ near threshold: the relative momentum of the quarks governing the
soft gluon exchange responsible for the Coulomb potential, and a high momentum
scale, induced by
hard gluon exchange, approximately equal to twice the quark mass for the
corrections.
\cite{Hoang} One thus can use threshold production to obtain a direct
determination of $\alpha_V$ even at low scales.  The corresponding QED results
for $\tau$ pair production allow for a measurement of the magnetic moment of the
$\tau$ and could be tested at a future $\tau$-charm
factory.\cite{Voloshin,Hoang}

\end{enumerate}

We also note that
computations in different sectors of the Standard Model have been
traditionally carried out using different renormalization schemes.
However, in a grand
unified theory, the forces between all of the particles in the fundamental
representation should become universal above the grand unification scale.
Thus it is
natural to use $\alpha_V$ as the effective charge for all sectors of a grand
unified theory,  rather than in
a convention-dependent coupling such as $\alpha_{\overline {MS}}$.
\section{The Analytic Extension of the $\bar{MS}$ Scheme}

The standard ${\overline {MS}}$ scheme
is not an analytic function of the renormalization scale at heavy quark
thresholds;
in the running of the coupling the quarks are taken as massless, and
at each quark threshold the value of $N_F$ which appears in the $\beta$
function is
incremented.  Thus Eq. (\ref{eq:csrmsovf}) is technically only valid far
above a heavy quark threshold.  However, we can use this commensurate scale
relation to define
an extended
$\overline {MS}$ scheme which is continuous and analytic at any scale.  The new
modified scheme inherits all of the good properties of the $\alpha_V$ scheme,
including its correct analytic properties as a function of the quark masses
and its
unambiguous scale fixing.\cite{bgmr}
Thus we define
\begin{equation}
\widetilde {\alpha}_{\overline{\mbox{\tiny MS}}}(Q)
= \alpha_V(Q^*) + \frac{2N_C}{3} {\alpha_V^2(Q^{**})\over\pi} +
\cdots ,
\label{alpmsbar2}
\end {equation}
for all scales $Q$.  This equation not only provides an analytic
extension of the $\overline{MS}$ and similar schemes, but it also ties down the
renormalization scale to the physical masses of the quarks as they
enter into the vacuum polarization contributions to $\alpha_V$.

The modified scheme \amst\ provides an analytic interpolation of
conventional $\overline{MS}$ expressions by utilizing the mass dependence of the
physical \av\ scheme.  In effect, quark thresholds are treated
analytically to all orders in $m^2/Q^2$; \ie, the evolution of the analytically
extended coupling in the intermediate regions reflects the actual mass
dependence of a physical effective charge and the analytic properties of
particle production.
Just as in Abelian QED, the mass dependence of the effective potential
and the analytically extended scheme \amst\ reflects the analyticity of the
physical thresholds for particle production in the
crossed channel.  Furthermore, the definiteness of the dependence in the quark
masses automatically constrains the renormalization scale.  There
is thus no scale ambiguity in perturbative expansions in \av\ or \amst.

In leading order the effective number of flavors in the modified scheme
\amst\ is given to a very good approximation by the simple form\cite{bgmr}
\begin{equation}
\widetilde {N}_{F,\overline{\mbox{\tiny MS}}}^{(0)}\left(\frac{m^2}{Q^2}\right)
\cong \left(1 + {5m^2 \over {Q^2\exp({5\over 3})}} \right)^{-1}
\cong \left( 1 + {m^2 \over Q^2} \right)^{-1}.
\end{equation}
Thus the contribution from one flavor is $\simeq 0.5$ when
the scale $Q$ equals the quark mass $m_i$.  The standard procedure
of matching $\alpha_{\overline{\mbox{\tiny MS}}}(\mu)$ at the quark
masses serves as a zeroth-order approximation to the continuous $N_F$.

\begin{figure}[htb]
\begin{center}
\leavevmode
\epsfxsize=4in
\epsfbox{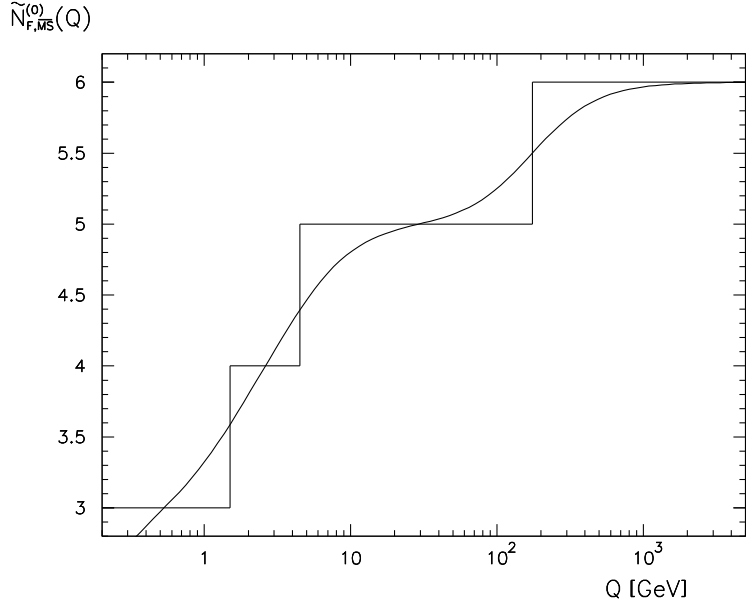}
\end{center}
\caption[*]{The continuous
$\widetilde {N}_{F,\overline{\mbox{\tiny MS}}}^{(0)}$ in the analytic
extension of the $\overline{\mbox{MS}}$ scheme as a
function of the physical scale $Q$.  (For reference the
continuous $N_F$ is also compared with
the conventional procedure of taking $N_F$ to be a step-function at the
quark-mass thresholds.)}
\label{fig:nfsum}
\end{figure}

Adding all flavors together gives the total
$\widetilde {N}_{F,\overline{\mbox{\tiny MS}}}^{(0)}(Q)$
which is shown in Fig.~\ref{fig:nfsum}.  For reference, the
continuous $N_F$ is also compared with
the conventional procedure of taking $N_F$ to be a step-function at the
quark-mass thresholds.
The figure shows clearly that there are hardly any plateaus at all
for the continuous
$\widetilde {N}_{F,\overline{\mbox{\tiny MS}}}^{(0)}(Q)$ in
between the quark masses.
Thus there is really no scale below 1 TeV where
$\widetilde {N}_{F,\overline{\mbox{\tiny MS}}}^{(0)}(Q)$
can be approximated by a constant; for all $Q$ below 1 TeV there is always
one quark
with mass $m_i$ such that $m_i^2 \ll Q^2$ or $Q^2 \gg m_i^2$ is not
true.
We also note that if one would use any other scale than the
BLM-scale for $\widetilde {N}_{F,\overline{\mbox{\tiny MS}}}^{(0)}(Q)$,
the result would be to increase the difference between the analytic
$N_F$ and the standard procedure of using the step-function at the
quark-mass thresholds.

\begin{figure}[htb]
\begin{center}
\leavevmode
\epsfxsize=4in
\epsfbox{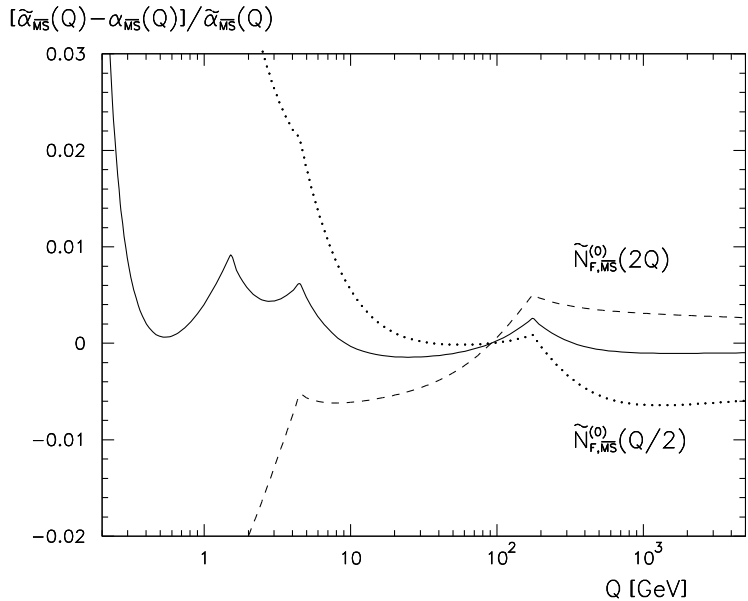}
\end{center}
\caption[*]{The solid curve shows the relative difference between the
solutions to
the 1-loop renormalization group equation using continuous $N_F$,
$\widetilde{\alpha}_{\overline{\mbox{\tiny MS}}}(Q)$, and conventional discrete
theta-function thresholds, $\alpha_{\overline{\mbox{\tiny MS}}}(Q)$.
The dashed (dotted) curves shows the same quantity but using the scale $2Q$
($Q/2$)
in $\widetilde {N}_{F,\overline{\mbox{\tiny MS}}}^{(0)}$.  The solutions
have been
obtained numerically starting from the world average\cite{Burrows}\
$\alpha_{\overline{\mbox{\tiny MS}}}(M_Z) = 0.118$.}
\label{fig:adiff}
\end{figure}

Figure~\ref{fig:adiff} shows the relative difference between the two
different solutions of the 1-loop renormalization group equation,
\ie\ $(\widetilde{\alpha}_{\overline{\mbox{\tiny MS}}}(Q)-
           {\alpha}_{\overline{\mbox{\tiny MS}}}(Q) )/
           \widetilde{\alpha}_{\overline{\mbox{\tiny MS}}}(Q)$.
The solutions have been obtained numerically starting from the
world average\cite{Burrows}
$\alpha_{\overline{\mbox{\tiny MS}}}(M_Z) = 0.118$.
The figure shows that
taking the quark masses into account in the running leads to
effects of the order of one percent which are most especially
pronounced near thresholds.

The extension
of the $\overline{\mbox{MS}}$-scheme proposed here provides a coupling
which is an analytic function of both the scale and the quark masses.
The new
modified coupling $\widetilde {\alpha}_{\overline{\mbox{\tiny MS}}}(Q)$
inherits most of the good properties of the $\alpha_V$ scheme, including its
correct analytic properties as a function of the quark masses and its
unambiguous scale fixing~\cite{bgmr}.
However, the conformal coefficients in the commensurate scale
relation between the $\alpha_V$ and
$\overline{\mbox{MS}}$ schemes does not preserve one of the
defining criterion
of the potential expressed in the bare charge, namely the non-occurrence of
color factors corresponding to an iteration of the potential.  This
is probably an effect of the breaking of conformal invariance by
the $\overline{\mbox{MS}}$ scheme.  The breaking of conformal symmetry
has also been observed when dimensional regularization is used as a
factorization scheme in both
exclusive\cite{Brodsky:1986ve,Frishman,Muller} and
inclusive\cite{Blumlein} reactions.  Thus, it does not turn out to be
possible to extend the modified scheme
${\widetilde \alpha}_{\overline{\mbox{\tiny MS}}}$ beyond leading order
without running into an intrinsic contradiction with conformal
symmetry.

\section{Application of Commensurate Scale Relations to the Hard QCD
Pomeron}

The observation of rapidly increasing structure functions in deep
inelastic scattering at small-$x_{bj}$ and the observation
of rapidly increasing diffractive processes such as $\gamma^* p \to \rho
p$ at high energies at HERA is in agreement with the expectations of the
BFKL\cite{BFKL} QCD high-energy limit.  The highest eigenvalue,
$\omega^{max}$, of the leading order (LO) BFKL equation\cite{BFKL} is
related to the intercept of the Pomeron which in turn governs
the high-energy asymptotics of the cross sections: $\sigma \sim
s^{\alpha_{I \negthinspace P}-1} = s^{\omega^{max}}$.
The BFKL Pomeron intercept in LO turns out to be rather large:
$\alpha_{I \negthinspace P} - 1 =\omega_L^{max} =
12 \, \ln2 \, ( \alpha_S/\pi ) \simeq 0.55 $ for
$\alpha_S=0.2$; hence, it is very important to know the next-to-leading order
(NLO) corrections.

Recently the NLO corrections to the BFKL resummation
of energy logarithms were calculated\cite{FL,CC98} by employing the
modified minimal subtraction scheme ($\overline{\mbox{MS}}$)
\cite{Bar78} to regulate the ultraviolet divergences with arbitrary
scale setting.  The NLO corrections\cite{FL,CC98}
to the highest eigenvalue of the BFKL equation turn out to be negative
and even larger than the LO contribution for $\alpha_s > 0.157$.
It is thus important to analyze the NLO BFKL resummation of
energy logarithms
\cite{FL,CC98} in physical renormalization schemes and apply the BLM-CSR
method.  In fact, as shown in a recent paper\cite{BBFKL}, the
reliability of QCD predictions for the intercept of the BFKL Pomeron at
NLO when evaluated using BLM scale setting
\cite{BLM} within non-Abelian
physical schemes, such as the momentum space
subtraction (MOM) scheme\cite{Cel79,Pas80}
or the $\Upsilon$-scheme based on $\Upsilon \rightarrow ggg$ decay,
is significantly improved compared to the $\overline{\mbox{MS}}$-scheme.

The renormalization scale ambiguity problem can be resolved if one can
optimize the choice of scales and renormalization schemes according to some
sensible criteria.  In the BLM optimal scale
setting\cite{BLM}, the renormalization scales are chosen such that
all vacuum polarization effects from the QCD $\beta$-function are resummed
into the running couplings.  The coefficients of the perturbative series are
thus identical to the perturbative coefficients of the corresponding
conformally invariant theory with $\beta=0$.

In the present case one can show that within the
V-scheme (or the $\overline{\mbox{MS}}$-scheme)
the BLM procedure does not change significantly the value of
the NLO coefficient $r(\nu)$.
This can be understood since the V-scheme, as well as
$\overline{\mbox{MS}}$-scheme, are adjusted primarily to
the case when in the LO there are dominant QED (Abelian) type
contributions, whereas in the BFKL case there are important LO
gluon-gluon (non-Abelian) interactions.
Thus one can choose for the BFKL case the MOM-scheme
\cite{Cel79,Pas80} or the $\Upsilon$-scheme based on
$\Upsilon \rightarrow ggg$ decay.

Adopting BLM scale setting, the NLO BFKL eigenvalue
in the MOM-scheme is
\begin{equation}
\omega_{BLM}^{MOM}(Q^{2},\nu) =
N_C \chi_{L} (\nu) \frac{\alpha_{MOM}(Q^{MOM \, 2}_{BLM})}{\pi}
\Biggl[1 +
r_{BLM}^{MOM} (\nu) \frac{\alpha_{MOM}(Q^{MOM \, 2}_{BLM})}{\pi} \Biggr] ,
\end{equation}
$$
r_{BLM}^{MOM} (\nu) = r_{MOM}^{conf} (\nu) \, .
$$

The $\beta$-dependent part of the $r_{MOM}(\nu)$ defines the
corresponding BLM optimal scale
\begin{equation}
Q^{MOM \, 2}_{BLM} (\nu) = Q^2 \exp
\Biggl[ - \frac{4 r_{MOM}^{\beta}(\nu)}{\beta_0} \Biggr]
= Q^2 \exp \Biggl[ \frac
{1}{2}\chi_L (\nu) - \frac{5}{3} + 2 \biggl(1+\frac{2}{3} I \biggr) \Biggr].
 \nonumber
\label{qblm}
\end{equation}
At $\nu=0$ we
have $Q^{MOM \, 2}_{BLM} (0) = Q^2 \bigl( 4 \exp [2(1+2 I /3)-5/3] \bigr) \simeq
Q^2  \, 127$. Note that $Q^{MOM \, 2}_{BLM}(\nu)$ contains a large factor,
$\exp [- 4 T_{MOM}^{\beta}/\beta_0 ] = \exp [2(1+2 I /3)] \simeq 168$, which
reflects a large kinematic difference between MOM- and
$\overline{\mbox{MS}}$- schemes\cite{Cel83,BLM}.

One of the striking features of this analysis is that the NLO value for
the intercept of the BFKL Pomeron, improved by the BLM procedure, has a
very weak dependence on the gluon virtuality $Q^2$.
This agrees with the conventional Regge-theory where
one expects an universal intercept of the Pomeron without any $Q^2$-dependence.
The minor $Q^2$-dependence obtained, on one side, provides near insensitivity
of the results to the precise value of $\Lambda$, and, on the other side, leads
to approximate scale and conformal invariance.  Thus one may use conformal
symmetry\cite{Lipatov97,Lipatov86} for the continuation of the present results
to the case $t \neq 0$.

The NLO corrections to the BFKL
equation for the QCD Pomeron thus become controllable
and meaningful provided one uses physical renormalization scales and
schemes relevant to non-Abelian gauge theory.  BLM optimal scale setting
automatically sets the appropriate physical renormalization scale by
absorbing the non-conformal $\beta$-dependent coefficients.  The
strong renormalization scheme dependence of the NLO corrections to BFKL
resummation then largely disappears.  This is in contrast to the unstable
NLO results obtained in the conventional $\overline{\mbox{MS}}$-scheme with
arbitrary choice of renormalization scale.
A striking feature of the NLO BFKL Pomeron intercept in
the BLM approach is its very weak $Q^2$-dependence, which provides
approximate conformal invariance.

The new results presented here open new windows for applications
of NLO BFKL resummation to high-energy phenomenology.

Recently the $L3$ collaboration at LEP{L3} has presented new results for
the virtual photon cross section $\sigma(\gamma^*(Q_A) \gamma^*(Q_b) \to
{\rm hadrons}$ using double tagged $e^+ e^- \to e^+ e^- {\rm hadrons}.$
This process provides a remarkably clean possible test of the perturbative
QCD pomeron since there are no initial hadrons.\cite{bhsprl}
The calculation of $\sigma (\gamma^* \gamma^*)$ and  is discussed in
detail in references~\cite{bhsprl}.  We note here some important features:

i) for large virtualities,  $\sigma (\gamma^* \gamma^*)$ the
longitudinal cross section $\sigma_{LL}$ dominates and scales like
$1/Q^2$, where $Q^2 \sim {\mbox {max}} \{ Q_A^2, Q_B^2\} $.  This is
characteristic of the perturbative QCD prediction.  Models based on
Regge factorization (which work well in the soft-interaction regime
dominating $\gamma \, \gamma$ scattering near the mass shell) would
predict a higher power in $1/Q$.

ii) $\sigma (\gamma^* \gamma^*)$ is affected by logarithmic
corrections in the energy $s$ to all orders in $\alpha_s$.  As a result
of the BFKL summation of these contributions, the cross section rises
like a power in $s$, $\sigma \propto s^\lambda$.  The Born
approximation to this result (that is, the ${\cal O} (\alpha_s^2) $
contribution, corresponding to single gluon exchange
gives a constant cross section, $\sigma_{Born} \propto s^0$.
A fit to photon-photon
sub-energy dependence measured by L3 at
$\sqrt s_{e^+e^-} = 91~{\rm GeV}$ and $<Q_A^2>=<Q_A^2>
= 3.5~{\rm GeV}^2$
gives
$\alpha_P-1 = 0.28 \pm 0.05$.  The L3 data at $\sqrt s_{e^+e^-} = 183~{\rm
GeV}$ and
$<Q_A^2>=<Q_A^2> = 14~{\rm GeV}^2,$ gives $\alpha_P-1 = 0.40 \pm 0.07$
which shows a rise of the virtual photon cross section much stronger
than single gluon or soft pomeron exchange, but it is compatible with the
expectations from the NLO scale- and scheme-fixed BFKL predictions.  It
will be crucial to measure the $Q_A^2$ and $Q_B^2$ scaling and
polarization dependence and compare with the detailed predictions
of PQCD\cite{bhsprl}.

\section{Summary on Commensurate Scale Relations}

Commensurate scale relations have a number of attractive properties:
\begin{enumerate}
\item
The ratio of physical scales $Q_A/Q_B$ which appears in commensurate scale
relations
reflects the relative position of physical thresholds, \ie\, quark
anti-quark pair
production.
\item
The functional dependence and perturbative expansion of the CSR are identical to
those of a conformal scale-invariant theory where $\beta_A(\alpha_A)=0$
and $\beta_B(\alpha_B)=0$.
\item
In the case of theories approaching fixed-point behavior
$\beta_A(\bar\alpha_A)=0$ and
$\beta_B(\bar\alpha_B)=0$, the commensurate scale relation relates both
the ratio of
fixed point couplings $\bar\alpha_A/\bar\alpha_B$, and the ratio of
scales as the fixed point is approached.
\item
Commensurate scale relations satisfy the Abelian correspondence principle
\cite{Brodsky:1997jk};
\ie\ the non-Abelian gauge theory prediction reduces to Abelian theory for
$N_C \to 0$ at
fixed $ C_F\alpha_s$ and fixed $N_F/C_F$.
\item
The perturbative expansion of a commensurate scale relation has the same
form as a
conformal theory, and thus has no
$n!$ renormalon growth arising from the $\beta$-function.\cite{Gardi}
It is an interesting conjecture whether the perturbative expansion relating
observables to observable are in fact free of all $n!$ growth.  The
generalized Crewther relation, where the commensurate relation's perturbative
expansion forms a geometric series to all orders, has convergent behavior.
\end{enumerate}

Virtually any perturbative QCD prediction can be written in the form of a
commensurate
scale relation, thus eliminating any uncertainty due to renormalization
scheme or scale
dependence.  Recently it has been shown\cite{Brodsky:1998ua} how the
commensurate scale relation between the radiative corrections to
$\tau$-lepton decay and
$R_{e^+e^-}(s)$
can be generalized and empirically tested for arbitrary $\tau$ mass and nearly
arbitrarily functional dependence of the $\tau$ weak decay matrix element.

An essential feature of the \av(Q) scheme is the absence of any
renormalization scale ambiguity, since $Q^2$ is,  by definition, the square of
the physical momentum transfer.  The \av\ scheme naturally takes into
account quark mass
thresholds,  which is of particular phenomenological importance to QCD
applications in the mass region close to threshold.
As we have seen, commensurate scale relations provide
an analytic extension of the conventional \ms\ scheme in which many of
the advantages of the \av\ scheme are inherited by the \amst\ scheme,
but only minimal changes have to be made.
Given the commensurate scale relation connecting \amst\ to \av\, expansions in
\amst\ are effectively expansions in \av\ to the given order in perturbation
theory at a corresponding commensurate scale.

The calculation of $\psi_V^{(1)}$, the two-loop
term in the Gell-Mann Low function for the $\alpha_V$ scheme,
with massive quarks gives for the first time a gauge invariant
and renormalization scheme independent two-loop result for the
effects of quarks masses in the running of the coupling.  Renormalization
scheme independence is achieved by using the pole mass definition for
the ``light" quarks which contribute to the scale dependence of
the static heavy quark potential.  Thus the pole mass and the $V$-scheme
are closely connected and have to be used in conjunction to give
reasonable results.

It is interesting that the effective number of flavors in the two-loop
coefficient of the Gell-Mann Low function in the $\alpha_V$ scheme,
$N_{F,V}^{(1)}$, becomes negative for intermediate values of
$Q/m$.  This feature can be
understood as anti-screening from the non-Abelian contributions and
should be contrasted with the QED case where the effective number of
flavors becomes larger than $1$ for intermediate $Q/m$.  For small
$Q/m$ the heavy quarks decouple explicitly as expected in a physical
scheme, and for large $Q/m$ the massless result is retained.

The analyticity of the $\alpha_V$ coupling can be utilized to
obtain predictions for any perturbatively calculable observables
including the
finite quark mass effects associated with the running of the
coupling.  By employing the commensurate scale relation method,
observables which have been calculated in the
$\overline{\mbox{MS}}$ scheme can be related to the analytic V-scheme
without any scale ambiguity.  The commensurate scale relations
provides the relation between the physical scales of two effective charges
where they pass through a common flavor threshold.  We also note the
utility of the \av\ effective charge in
supersymmetric and grand unified theories, particularly since the
unification of couplings and masses would be expected to occur in terms
of physical quantities rather than parameters defined by theoretical
convention.

As an example we have showed in Ref. \cite{bgmr} how to calculate the finite quark mass
corrections connected with the running of the coupling for
the non-singlet hadronic width of the Z-boson compared
with the standard treatment in the $\overline{\mbox{MS}}$ scheme.
The analytic treatment in the V-scheme gives a simple and straightforward
way of incorporating these effects for any observable.  This should
be contrasted with the $\overline{\mbox{MS}}$ scheme where higher
twist corrections due to finite quark mass threshold effects have to be
calculated separately for each observable.
The V-scheme is especially
suitable for problems where the quark masses are important such as for
threshold production of heavy quarks and the hadronic width of the
$\tau$ lepton.

It has also been shown that the NLO corrections to
the BFKL equation for the QCD Pomeron become controllable
and meaningful provided one uses physical renormalization scales and
schemes relevant to non-Abelian gauge theory.  BLM optimal scale setting
automatically sets the appropriate physical renormalization scale by
absorbing the non-conformal $\beta$-dependent coefficients.  The
strong renormalization scheme dependence of the NLO corrections to BFKL
resummation then largely disappears.  This is in contrast to the unstable
NLO results obtained in the conventional $\overline{\mbox{MS}}$-scheme with
arbitrary choice of renormalization scale.
A striking feature of the NLO BFKL Pomeron intercept in
the BLM/CSR approach is its very weak $Q^2$-dependence, which provides
approximate conformal invariance.
The new results presented here open new windows for applications
of NLO BFKL resummation to high-energy phenomenology, particularly virtual
photon-photon scattering.

\begin{center}
{\bf Acknowledgments}
\end{center}

Many of the results presented here are based on collaborations with a
number of colleagues, including 
Victor Fadin, 
Gregory Gabadadze, 
Mandeep Gill, 
John Hiller, 
Dae Sung Hwang, 
Chueng Ji, 
Andrei Kataev, 
Victor Kim, 
Peter Lepage, 
Lev Lipatov, 
Hung Jung Lu, 
Gary McCartor,
Michael Melles, 
Chris Pauli, 
Grigorii B. Pivovarov. 
and 
Johan Rathsman. 
I thank
S. Dalley, 
Yitzhak Frishman, 
Einan Gardi, 
Georges Grunberg, 
Paul Hoyer, 
Marek Karliner, 
Carlos Merino, 
and 
Jose Pelaez
for helpful
conversations. The sections on commensurate scale relations are based on
a review written in collaboration with Johan
Rathsman.\cite{Brodsky:1999gm} I also wish to thank Chueng Ji and
Dong-Pil Min for their outstanding hospitality at the Asia Pacific Center
for Theoretical Physics in Seoul.


\newpage

\begin{thebibliography}{99}

\bibitem{Nasalski:1994bh}
J.P.~Nasalski [New Muon Collaboration],
Nucl.\ Phys.\ {\bf A577}, 325C (1994).

\bibitem{Barone:1999yv}
V.~Barone, C.~Pascaud and F.~Zomer, hep-ph/9907512.

\bibitem{Karliner:1999fn}
M.~Karliner and H.J.~Lipkin, hep-ph/9906321.

\bibitem{LB}
G. P. Lepage and S. J. Brodsky, {\em Phys. Rev.} {\bf D22}, 2157 (1980);
{\em Phys. Lett.} {\bf B87}, 359 (1979); {\em Phys. Rev. Lett.} {\bf 43},
545, 1625(E) (1979).

\bibitem{PinskyPauli}
S. J.~Brodsky, H.~Pauli and S. S.~Pinsky,
{\em Phys. Rept.}  {\bf 301}, 299 (1998) hep-ph/9705477.

\bibitem{Bertsch}
G. Bertsch, S. J. Brodsky,
A. S. Goldhaber, and J. F. Gunion, {\em Phys. Rev. Lett.} {\bf 47}, 297 (1981).

\bibitem{MillerFrankfurtStrikman}
L. Frankfurt, G. A. Miller, and M. Strikman, {\em Phys. Lett.}
{\bf B304}, 1 (1993),  hep-ph/9305228.

\bibitem{BD}
S. J. Brodsky and S. D. Drell, {\em Phys. Rev.} {\bf D22}, 2236 (1980).

\bibitem{BrodskyLepage}
S. J. Brodsky and G. P. Lepage, in {\em Perturbative Quantum
Chromodynamics}, A. H. Mueller, Ed.  (World Scientific, 1989).

\bibitem{DLCQ}
H. C. Pauli and S. J. Brodsky, {\em Phys. Rev.} {\bf D32}, 1993 (1985);
{\em Phys. Rev.} {\bf D32}, 2001 (1985).

\bibitem{Brodsky:1991ir}
S.J.~Brodsky and H.C.~Pauli,
{\it lectures given at 30th Schladming Winter School in Particle
                  Physics: Field Theory, Schladming, Austria, Feb 27 - Mar 8,
                  1991}.

\bibitem{Kleb}
S. Dalley, and I. R. Klebanov, {\em Phys. Rev.} {\bf D47}, 2517 (1993).

\bibitem{AD}
F. Antonuccio and S. Dalley, {\em Phys. Lett.} {\bf B348}, 55 (1995);
{\em Phys. Lett.} {\bf B376}, 154 (1996); {\em Nucl. Phys.} {\bf B461}, 275
(1996).

\bibitem{Brodsky:1998hs}
S.J.~Brodsky, J.R.~Hiller and G.~McCartor,
Phys.\ Rev.\ {\bf D58}, 025005 (1998)
hep-th/9802120.

\bibitem{Srivastava:1999gi}
P.P.~Srivastava and S.J.~Brodsky, hep-ph/9906423.


\bibitem{BLM}
S. J.~Brodsky, G. P.~Lepage and P. B.~Mackenzie,
Phys. Rev. {\bf D28}, 228 (1983).

\bibitem{CSR}
S.J.~Brodsky, H.J.~Lu, Phys.~Rev.~{\bf D51}, 3652 (1995); hep-ph/9506322.

\bibitem{BrodskyKataevGabaladzeLu}
S. J.  Brodsky, G. T. Gabadadze, A. L. Kataev and H. J. Lu,
{\em Phys. Lett.} {\bf 372B}, 133, (1996).

\bibitem{BJPR}
S. J. Brodsky, C.-R. Ji, A. Peng and D. G. Robertson,
Phys. Rev. {\bf D57}, 345 (1998).

\bibitem{Watson:1997fg}
N.J.~Watson,
Nucl.\ Phys.\ {\bf B494}, 388 (1997)
hep-ph/9606381.

\bibitem{Czarnecki:1998sz}
A.~Czarnecki, K.~Melnikov and N.~Uraltsev,
Phys.\ Rev.\ Lett.\ {\bf 80}, 3189 (1998)
hep-ph/9708372.

\bibitem{Brodsky:1998hn}
S.J.~Brodsky and D.S.~Hwang,
Nucl.\ Phys.\ {\bf B543}, 239 (1999)
hep-ph/9806358.

\bibitem{CRY}
S. J. Chang, R.G. Root and T. M. Yan, {\em Phys. Rev.} {\bf D7}, 1133 (1973).

\bibitem{BUR}
M. Burkardt, {\em Nucl. Phys.} {\bf A504}, 762 (1989);
{\em Nucl. Phys.} {\bf B373}, 613 (1992); {\em Phys. Rev.} {\bf D52}, 3841
(1995).

\bibitem{Choi:1998nf}
H.~Choi and C.~Ji,
Phys.\ Rev.\ {\bf D58}, 071901 (1998)
hep-ph/9805438.

\bibitem{Horn}
K.  Hornbostel, S. J. Brodsky, and H. C.  Pauli,
{\em Phys. Rev.} {\bf D41} 3814 (1990).

\bibitem{BHS}
A. Szczepaniak, E. M. Henley and S. J. Brodsky,
{\em Phys. Lett.} {\bf B243}, 287 (1990).

\bibitem{Sz}
A.  Szczepaniak, {\em Phys. Rev.} {\bf D54}, 1167 (1996).

\bibitem{BALL}
P.~Ball,
JHEP {\bf 09}, 005 (1998)
hep-ph/9802394.

\bibitem{BABR}
P.~Ball and V. M.~Braun, {\em Phys. Rev.} {\bf D58}, 094016 (1998)
hep-ph/9805422.


\bibitem{BGMFS}
S. J. Brodsky, L. Frankfurt, J. F. Gunion, A. H.
Mueller, and M. Strikman,  {\em Phys. Rev.} {\bf D50}, 3134 (1994),
hep-ph/9402283.

\bibitem{BM}
S. J. Brodsky and A. H. Mueller, {\em Phys. Lett.} {\bf 206B}, 685 (1988).
  L. Frankfurt and M. Strikman, {\em Phys. Rept.} {\bf  160} , {235} (1988);
  P. Jain, B. Pire and J. P. Ralston, {\em Phys. Rept.} {\bf 271}, {67}( 1996).

\bibitem{ABD}
F. Antonuccio,  S. J. Brodsky, and S. Dalley, SLAC-PUB-7472,
{\em Phys. Lett.} {\bf B412} 104 (1997), hep-ph/9705413.

\bibitem{Mueller}
A.  H.  Mueller,  Phys. Lett.  {\bf B308}, 355 (1993).

\bibitem{Dmuller}
D. Mueller, SLAC-PUB-6496, May 1994,   hep-ph/9406260.

\bibitem{IC}
S. J. Brodsky, P. Hoyer, C. Peterson, and N. Sakai, {\em Phys. Lett.} {\bf
93B},  451 (1980).

\bibitem{HSV}
B. W. Harris,  J. Smith, and R. Vogt,  {\em Nucl. Phys.} {\bf B461}, 181 (1996),
hep-ph/9508403.

\bibitem{BS}
S. J. Brodsky and I. A. Schmidt, {\em Phys. Lett.} {\bf B234}, 144 (1990).

\bibitem{BHMT}
S. J. Brodsky, P. Hoyer, A. H. Mueller, W.-K. Tang, {\em Nucl. Phys.} {\bf
B369}, 519 (1992).

\bibitem{Brodsky:1997fj}
S. J.~Brodsky and M.~Karliner,
{\em Phys. Rev. Lett.} {\bf 78}, 4682 (1997)
hep-ph/9704379.

\bibitem{Warr}
M. Burkardt and  Brian Warr, {\em Phys. Rev.} {\bf D45}, 958 (1992).

\bibitem{Signal}
A. I. Signal and A. W. Thomas, {\em Phys. Lett.} {\bf 191B}, 205 (1987).

\bibitem{BMa}
S. J. Brodsky and B-Q Ma, {\em Phys. Lett.} {\bf B381}, 317 (1996),
hep-ph/9604393.

\bibitem{BrodskyMueller}
S. J. Brodsky and A. Mueller, {\em Phys. Lett.} {\bf 206B}, 685 (1988).
R. Vogt, S. J. Brodsky, and P. Hoyer,
SLAC-PUB-5421,{\em  Nucl. Phys.} {\bf B360}, 67 (1991);
SLAC-PUB-5827, {\em  Nucl. Phys.} {\bf B383}, 643 (1992).

\bibitem{Hoyer:1998ha}
P.~Hoyer and S.~Peigne,
Phys.\ Rev.\ {\bf D59}, 034011 (1999)
hep-ph/9806424.

\bibitem{BrodskyBerger}
E. L. Berger and S. J. Brodsky, {\em  Phys. Rev. Lett.} {\bf 42}, 940 (1979).

\bibitem{Brandenburg}
A. Brandenburg, S. J. Brodsky, V.V. Khoze, and D. Mueller, {\em Phys. Rev.
Lett.} {\bf 73}, 939 (1994), hep-ph/9403361.

\bibitem{Brodsky:1976rz}
S. J.~Brodsky and B. T.~Chertok, {\em Phys. Rev.} {\bf D14}, 3003 (1976).

\bibitem{bjl83}
S. J. Brodsky, C.-R. Ji, and G. P. Lepage, {\em Phys. Rev. Lett.} {\bf
51}, 83 (1983).

\bibitem{Farrar:1991qi}
G. R.~Farrar, K.~Huleihel and H.~Zhang,
{\em Phys. Rev. Lett.} {\bf 74}, 650 (1995).

\bibitem{krisch92}
A. D. Krisch, {\em Nucl. Phys. B (Proc. Suppl.)} {\bf 25}, 285 (1992).

\bibitem{Brodsky:1988xw}
S. J.~Brodsky and G. F.~de Teramond,
{\em Phys. Rev. Lett.} {\bf 60}, 1924 (1988).

\bibitem{Gronberg:1998fj}
J.~Gronberg {\it et al.} [CLEO Collaboration],
{\em Phys. Rev.} {\bf D57}, 33 (1998)
hep-ex/9707031.

\bibitem{E791}
D. F. Ashery \etal,  Fermilab E791 Collaboration, to be published.

\bibitem{Beneke:1999br}
M.~Beneke, G.~Buchalla, M.~Neubert and C.T.~Sachrajda,
hep-ph/9905312.

\bibitem{Brodsky:1980ny}
S. J.~Brodsky, Y.~Frishman, G. P.~Lepage and C.~Sachrajda,
{\em Phys. Lett.} {\bf 91B}, 239 (1980).

\bibitem{Brodsky:1986ve}
S. J.~Brodsky, Y.~Frishman and G. P.~Lepage,
{\em Phys. Lett.} {\bf 167B}, 347 (1986).

\bibitem{Muller:1994hg}
D.~M\"uller, {\em Phys. Rev.} {\bf D49}, 2525 (1994).

\bibitem{Ball:1998ff}
P.~Ball and V. M.~Braun, {\em Nucl. Phys.} {\bf B543}, 201 (1999)
hep-ph/9810475.

\bibitem{Braun:1999te}
V.M.~Braun, S.E.~Derkachov, G.P.~Korchemsky and A.N.~Manashov,
hep-ph/9902375.

\bibitem{Brodsky:1981kj}
S. J.~Brodsky and G. P.~Lepage,
{\em Phys. Rev.} {\bf D24}, 2848 (1981).

\bibitem{Frankfurt:1992dx}
L.~Frankfurt, G. A.~Miller and M.~Strikman,
{\em Comments Nucl. Part. Phys.} {\bf 21}, 1 (1992).

\bibitem{Brodsky:1995eh}
S. J.~Brodsky and H. J.~Lu,
{\em Phys. Rev.} {\bf D51}, 3652 (1995)
hep-ph/9405218.

\bibitem{Brodsky:1996tb}
S. J.~Brodsky, G. T.~Gabadadze, A. L.~Kataev and H. J.~Lu,
{\em Phys. Lett.} {\bf B372}, 133 (1996) hep-ph/9512367.

\bibitem{Brodsky:1999gm}
S. J.~Brodsky and J.~Rathsman, hep-ph/9906339.

\bibitem{Brodsky:1998dh}
S.J.~Brodsky, C.~Ji, A.~Pang and D.G.~Robertson,
Phys.\ Rev.\ {\bf D57}, 245 (1998)
hep-ph/9705221.

\bibitem{BF}
S. J.~Brodsky and G. R.~Farrar,
{\em Phys. Rev.} {\bf D11}, 1309 (1975).

\bibitem{Matveev:1973ra}
V. A.~Matveev, R. M.~Muradian and A. N.~Tavkhelidze,
{\em Nuovo Cim. Lett.} {\bf 7}, 719 (1973).

\bibitem{Collins:1997fb}
J.C.~Collins, L.~Frankfurt and M.~Strikman,
Phys.\ Rev.\ {\bf D56}, 2982 (1997)
hep-ph/9611433.

\bibitem{acw}
  C. E. Carlson and A. B. Wakely, {\em Phys. Rev.} {\bf D48}, {2000} (1993);
  A. Afanasev, C. E. Carlson and C. Wahlquist, {\em Phys. Lett.} {\bf B398},
{393} (1997),
  hep-ph/9701215, and {\em Phys. Rev.} {\bf D58}, {054007} (1998),
hep-ph/9706522.

\bibitem{Brodsky:1998sr}
S. J.~Brodsky, M.~Diehl, P.~Hoyer and S.~Peigne,
{\em Phys. Lett.} {\bf B449}, 306 (1999) hep-ph/9812277.

\bibitem{BB}
  J. F. Gunion, S. J. Brodsky and R. Blankenbecler, {\em Phys. Rev.} {\bf
D6}, {2652} (1972);
  R. Blankenbecler and S. J. Brodsky, {\em Phys. Rev.} {\bf D10}, {2973} (1974).

\bibitem{Landshoff:1974ew}
P.V.~Landshoff, {\em Phys. Rev.} {\bf D10}, 1024 (1974).

\bibitem{Chernyak:1999cj}
V.~Chernyak, hep-ph/9906387.

\bibitem{Adams:1997bh}
M.R.~Adams {\it et al.} [E665 Collaboration],
Z.\ Phys.\ {\bf C74}, 237 (1997).

\bibitem{Frankfurt:1999tq}
L.~Frankfurt, G. A.~Miller and M.~Strikman,
hep-ph/9907214.

\bibitem{Kroll}
P. Kroll and M. Raulfs, {\em Phys. Lett.} {\bf B387}, 848 (1996).

\bibitem {Rad}
I. V. Musatov and A. V. Radyushkin,  {\em Phys. Rev.} {\bf D56},
2713 (1997).

\bibitem{Feldmann:1999wr}
T.~Feldmann, hep-ph/9907226.

\bibitem{Schmedding:1999ap}
A.~Schmedding and O.~Yakovlev, hep-ph/9905392.

\bibitem{Stoler:1999nj}
P.~Stoler,
Few Body Syst.\ Suppl.\ {\bf 11}, 124 (1999).

\bibitem{Holt:1990ze}
R. J.~Holt, {\em Phys. Rev.} {\bf C41}, 2400 (1990).

\bibitem{Bochna:1998ca}
C.~Bochna {\it et al.} [E89-012 Collaboration],
{\em Phys. Rev. Lett.} {\bf 81}, 4576 (1998)
nucl-ex/9808001.

\bibitem{Melic:1999hg}
B.~Melic, B.~Nizic and K.~Passek, hep-ph/9903426.

\bibitem{Szczepaniak:1998sa}
A.~Szczepaniak, A.~Radyushkin and C.~Ji,
{\em Phys. Rev.} {\bf D57}, 2813 (1998)
hep-ph/9708237.

\bibitem{Bebek:1976ww}
C. J.~Bebek {\it et al.}, {\em Phys. Rev.} {\bf D13}, 25 (1976).

\bibitem{Isgur:1989iw}
N.~Isgur and C. H.~Llewellyn Smith,
{\em Phys. Lett.} {\bf B217}, 535 (1989).

\bibitem{Radyushkin:1998rt}
A. V.~Radyushkin, {\em Phys. Rev.} {\bf D58}, 114008 (1998)
hep-ph/9803316.

\bibitem{Bolz:1996sw}
J.~Bolz and P.~Kroll, {\em Z. Phys.} {\bf A356}, 327 (1996)
hep-ph/9603289.

\bibitem{Kronfeld:1991kp}
A.S.~Kronfeld and B.~Nizic,
Phys.\ Rev.\ {\bf D44}, 3445 (1991).

\bibitem{Brodsky:1972vv}
S. J.~Brodsky, F. E.~Close and J. F.~Gunion,
{\em Phys. Rev.} {\bf D6}, 177 (1972).

\bibitem{Brodsky:1989pv}
For reviews and further references see S. J.~Brodsky and G. P.~Lepage,
SLAC-PUB-4947.  Published in 'Perturbative Quantum Chromodynamics', Ed. by A.H.
Mueller,  World Scientific Publ. Co.  (1989), p. 93-240 (QCD161:M83);
V. L.~Chernyak and A. R.~Zhitnitsky,
{\em Phys. Rept.} {\bf 112}, 173 (1984).

\bibitem{Brodsky:1997jk}
S.J.~Brodsky and P.~Huet, Phys. Lett. {\bf B417}, 145 (1998)
hep-ph/9707543.

\bibitem{JiRad}
  X. Ji, {\em Phys. Rev.} {\bf D55}, {7114} (1997), hep-ph/9609381;
  X. Ji and J. Osborne, {\em Phys. Rev.}{\bf D58}, {094018} (1998),
hep-ph/9801260;
  A.V. Radyushkin, {\em Phys. Rev.}{\bf D56}, {5524} (1997), hep-ph/9704207.

\bibitem {BGKL}
S.  J. Brodsky , G. T. Gabadadze,  A. L. Kataev, and H. J. Lu.
Phys. Lett. {\bf B 372}, 133 (1996).

\bibitem{Grunberg}
G. Grunberg, Phys. Lett. {\bf B85}. 70 (1980);
Phys. Lett. {\bf B110}, 501 (1982);
Phys. Rev. {\bf D29}, 2315 (1984).

\bibitem{DharGupta}
A. Dhar and V. Gupta, Phys. Rev. {\bf D29}, 2822 (1984).

\bibitem{GuptaShirkovTarasov}
V. Gupta, D. V. Shirkov and O. V. Tarasov,
Int. J. Mod. Phys.  {\bf A6}, 3381  (1991).

\bibitem{Uraltsev}
A. Czarnecki, K. Melnikov, and N. Uraltsev,
Phys. Rev. Lett. {\bf 80}, 3189 (1998).
Yu. L. Dokshitser  and  B. R. Webber,  Phys. Lett. {\bf B404}, 321 (1997).

\bibitem{Adler}
S. Adler,  Phys. Rev.  {\bf 182}, 1517  (1969).

\bibitem{MattinglyStevenson}
A. C. Mattingly and P. M. Stevenson, Phys. Rev. {\bf D49}, 437 (1994).

\bibitem{CCFRL1}
CCFR Collaboration, W.C. Leung, {\it et al.},  Phys. Lett. {\bf B317}, 655
(1993).

\bibitem{CCFRL2}
CCFR and NuTeV Collaboration, presented by D. Harris at XXX Recontre de
Moriond, 1995,  presented by J. H. Kim at the  European
Conference on High Energy Physics, Brussels, July 1995.

\bibitem{KS}
A. L. Kataev, A.V. Sidorov,  Phys. Lett. {\bf  B331}, 179 (1994).

\bibitem{CCFRQ}
CCFR Collaboration, P.Z. Quinta, {\it et al.},  Phys. Rev. Lett. {\bf
71}, 1307 (1993).

\bibitem{tHooft}
G. 't Hooft, in the
Proceedings of the International School, Erice, Italy, 1977, edited by A.
Zichichi, Subnuclear Series Vol. 15 (Plenum, New York, 1979).

\bibitem{LuOneDim}
H. J. Lu, Phys. Rev. {\bf D45}, 1217 (1992)..

\bibitem{BenekeBraun}
M. Beneke  and  V. M. Braun, Phys. Lett. {\bf B348}, 513 (1995).


\bibitem{LepageMackenzie}
G. Peter Lepage and  P.  B. Mackenzie,
Phys. Rev. {\bf D48}, 2250 (1993).

\bibitem{Neubert}
M.  Neubert,  Phys. Rev. {\bf D51}, 5924 (1995).

\bibitem{BallBenekeBraun}
P. Ball, M. Beneke  and  V. M. Braun,
Nucl. Phys. {\bf B452}, 563 (1995).

\bibitem{derujula}
A.~De R{\'u}jula and H.~Georgi, Phys.~Rev.~{\bf D13}, 1296 (1976).

\bibitem{Georgi_Politzer}
H.~Georgi and H.D.~Politzer, Phys.~Rev.~{\bf D14}, 1829 (1976).

\bibitem{shirkov}
D.~V.~Shirkov, Teor.~Mat.~Fiz. {\bf 98}, 500 (1992)
[Theor. Math. Phys. {\bf 93}, 1403 (1992)];
D.~V.~Shirkov and S.~V.~Mikhailov, Zeit.~Phys.~{\bf C63}, 463 (1994).

\bibitem{chyla}
J.~Ch{\'y}la, Phys.~Lett.~{\bf B 351}, 325 (1995).

\bibitem{ac} T.~Appelquist, J.~Carazzone,
Phys.~Rev.~{\bf D11}, 2856 (1975).

\bibitem{melles98}
M.~Melles,
hep-ph/9805216, Phys.~Rev.~{\bf D58}:114004, 1998.

\bibitem{vegas} G.P.~Lepage, J.~Comp.~Phys.~{\bf 27}, 192 (1978);
G.P.~Lepage, Cornell preprint, CLNS-80/447, March 1980.

\bibitem{Peter}
M. Peter,  Nucl. Phys. {\bf B501}, 471 (1997).

\bibitem{bgmr}
S. J. Brodsky, M. S. Gill, M. Melles,  and J.  Rathsman,
Phys. Rev. {\bf D58}, 116006 (1998).

\bibitem{yh} T.~Yoshino, K.~Hagiwara,
Z.Phys. {\bf C 24}, 185 (1984).

\bibitem{jt} F.~Jegerlehner, O.V.~Tarasov,
hep-ph/9809485 and DESY 98-093.

\bibitem{Davies}
C.T.H.~Davies, {\it et al.}, Phys. Lett. {\bf B345}, 42 (1995);
Phys. Rev. {\bf D56}, 2755 (1997).

\bibitem{Henley}
A. Szczepaniak, E. M. Henley, S. J. Brodsky,
 Phys. Lett. {\bf B243}, 287  (1990).

\bibitem{Voloshin}
 B.  H. Smith, M. B. Voloshin,  Phys. Lett. {\bf B324}, 117  (1994).
Erratum-ibid. {\bf B333}, 564 (1994).

\bibitem{Hoang}
S.J. Brodsky, A. H. Hoang, J. H. Kuhn,  and T. Teubner,
Phys. Lett.  {\bf B359}, 355 (1995).

\bibitem {Fadin}
V. S. Fadin, V. A. Khoze, A. D. Martin, and W. J. Stirling,
 Phys. Lett.  {\bf B363}, 112  (1995).

\bibitem{Burrows}
P. N.~Burrows, Acta Phys.~Polon.~{\bf B28}, 701 (1997).

\bibitem{Frishman}
S. J.~Brodsky, P.~Damgaard, Y.~Frishman and G. P.~Lepage,
Phys. Rev. {\bf D33}, 1881 (1986).

\bibitem{Muller}
D.~Muller, Phys. Rev. {\bf D59}, 116001 (1999); A. V.~Belitsky and D.~Muller,
Nucl. Phys. {\bf B537}, 397 (1999); D.~Muller, Phys. Rev. {\bf D49}, 2525
(1994).

\bibitem{Blumlein}
J.~Blumlein, V.~Ravindran and W. L.~van Neerven, hep-ph/9812450.

\bibitem{BFKL}
V.~S.~Fadin, E.~A.~Kuraev and L.~N.~Lipatov,
Phys. Lett. {\bf 60B},  50 (1975);
L.~N.~Lipatov, Yad. Fiz. {\bf 23}, 642 (1976)
[Sov. J. Nucl. Phys. {\bf 23}, 338 (1976)];
E.~A.~Kuraev, L.~N.~Lipatov and V.~S.~Fadin, Zh. Eksp. Teor. Fiz.
{\bf 71}, 840 (1976) [Sov. JETP {\bf 44}, 443 (1976)];
{\it ibid.} {\bf 72}, 377 (1977) [{\bf 45}, 199 (1977)];
Ya.~Ya.~Balitski\v i and L.~N.~Lipatov, Yad. Fiz. {\bf 28}, 1597 (1978)
[Sov. J. Nucl. Phys. {\bf 28}, 822 (1978)].

\bibitem{FL}
V.~S.~Fadin and L.~N.~Lipatov, Phys. Lett. {\bf 429B}, 127
(1998).

\bibitem{CC98}
G.~Camici and M.~Ciafaloni, Phys. Lett. {\bf 430B}, 349
(1998).

\bibitem{Bar78}
W.~A.~Bardeen, A.~J.~Buras, D.~W.~Duke and T.~Muta,
Phys.~Rev.  {\bf  D18}, 3998 (1978).

\bibitem{BBFKL}
S.J.~Brodsky, V.S.~Fadin, V.T.~Kim, L.N.~Lipatov and G.B.~Pivovarov,
hep-ph/9901229.

\bibitem{Cel79}
W.~Celmaster and R.~J.~Gonsalves,
Phys.~Rev. {\bf D20}, 1420 (1979); \newline
Phys.~Rev.~Lett.~{\bf 42}, 1435 (1979).

\bibitem{Pas80}
P.~Pascual and R.~Tarrach, Nucl.~Phys.~{\bf B174}, 123
(1980); (E) {\bf B181}, 546 (1981).

\bibitem{Cel83}
W.~Celmaster and P.~M.~Stevenson,
Phys.~Lett.~{\bf 125B}, 493 (1983).

\bibitem{Lipatov97}
L.~N.~Lipatov, Phys.~Rept. {\bf C286}, 131 (1997).

\bibitem{Lipatov86}
L.~N.~Lipatov, Zh.~Eksp.~Teor.~Fiz. {\bf 90}, 1536 (1986)
[Sov. JETP {\bf 63}, 904 (1986)];
in {\it Perturbative Quantum Chromodynamics}, ed. A.H. Mueller
(World Scientific, Singapore, 1989) p. 411;
R.~Kirschner and L.~Lipatov, Zeit.~Phys.~{\bf C45}, 477 (1990).

\bibitem{bhsprl}
      S. J.~Brodsky, F.~Hautmann and D.E.~Soper,
      Phys. Rev. {\bf D56}, 6957 (1997);
      S. J.\ Brodsky, F.\ Hautmann and D. E.\ Soper, Phys.\ Rev.\ Lett.\
      {\bf 78}, 803 (1997);
      F.\ Hautmann,   talk at ICHEP96 (Warsaw, July 1996),
      preprint OITS 613/96,       in Proceedings of the
       XXVIII International Conference on High Energy Physics,
      eds. Z.\ Ajduk and A. K.\ Wroblewski, World Scientific, p.705;
      J.\ Bartels, A.\ De Roeck and H.\ Lotter,
      Phys.\ Lett. {\bf B389}, 742 (1996).

\bibitem{Gardi} S. J. Brodsky, E. Gardi, G. Grunberg, and J. Rathsman (in
preparation).

\bibitem{Brodsky:1998ua}
S.J.~Brodsky, J.R.~Pelaez and N.~Toumbas,
Phys.\ Rev.\ {\bf D60}, 037501 (1999)
hep-ph/9810424.

\end{thebibliography}
\end{document}